\documentclass[a4paper]{article}

\PassOptionsToPackage{pdftex,breaklinks, colorlinks=true, citecolor=blue}{hyperref}

\usepackage[latin1]{inputenc}
\usepackage[shortlabels]{enumitem}

\usepackage{siunitx}
\usepackage{amsmath}
\usepackage{graphicx}

\newtheorem{teo}{\sc Theorem}
\newtheorem{defi}[teo]{\sc Definition}

\usepackage[nosort]{cite}

\begin{document}

\title{A theoretical model for realistic local climates}
\author{Gabriele Di Bona\footnote{Queen Mary University of London, School of Mathematical Science, Complex Systems and Networks, Mile End Road, London E1 4NS, United Kingdom, g.dibona@qmul.ac.uk} \footnote{Scuola Superiore di Catania, Via Valdisavoia 9, 95123 Catania, Italy} \and Andrea Giacobbe\footnote{Università degli Studi di Catania, Dipartimento di Matematica e Informatica, Viale Andrea Doria 6, 95125 Catania, Italy, giacobbe@dmi.unict.it}}
\date{Jan 23, 2020}
\maketitle


\begin{abstract}
\noindent We write a nonlinear model that predicts the climate (temperature and humidity) on the surface of a small region on Earth, perform numerical investigations using the model, and compare the results to real climate on a variety of regions on Earth. It the parameters are chosen keeping into consideration the climatic K\"oppen zone to which the region belongs, the numerical model accurately reproduces the real climate.

\noindent The model takes into account the doubly-periodic forcing of the solar radiation (annual and daily), the laws of irradiance, the fact that the Earth has land and oceans with different thermic inertia, and the humidity of the air due to evaporation. This enables us to reproduce remarkable features of Earth's climate such as lag of seasons, lag of noons, and asymmetric evolution of daily temperatures.

\noindent The model can easily be adapted to planets with non-terrestrial astronomic parameters. We conclude this article with an investigation of an Earth with eccentricity higher than real.


\end{abstract}

\section{Introduction}

Since the origin of modern meteorology, in the late 1800, researchers have based their models on a balance between incoming and outgoing radiative energy. Meteorology is in fact driven by such energy, and its main macroscopic indicator, the temperature, depends on the thermal inertia of the materials being irradiated and on the heat exchange among the different materials composing the surface of the planet. Relying on these ingredients, early meteorologists considered weather forecasting impossible \cite{2003.WEA.Lynch}.

Rapid local variations of temperature and other measurable quantities are due to motion of fluid masses and, excluding systematic effects due to Coriolis force and conformation of oceans bed and shores, such variations reasonably average out. It follows that the evolution of average macroscopic thermodynamic quantities is reasonably not as much influenced by such phenomena \cite{2010.McGuffie.Henderson-Sellers}. This is the main difference between \emph{meteorology} and \emph{climate}. For the reasons above, many aspects of climate can, and maybe should, be investigated disregarding meteorological models. In particular, many climatic effects can be exposed reconsidering Margules and Richardson's basic models which, despite their simplicity, can give accurate climatic predictions. 

In this article we focus on three main climatic effects. The first phenomenon is called \emph{lag of seasons}. With this name one indicates the well known fact that the warmer days of the year take place some time after the days of maximal solar irradiance. The second phenomenon is the less celebrated phenomenon of \emph{lag of noons}. With this name we indicate the fact that the warmer hours of the day take place some time after the hours of strongest solar irradiance. The third phenomenon is the \emph{asymmetries of temperatures}. With this name we indicate the fact that daily temperatures rise much faster in the morning than they fall in the afternoon; this happens despite the fact that the forcing term is perfectly symmetric in shape, and an approach that takes into consideration only sun's irradiation and Fourier's law cannot reproduce this phenomenon. To obtain a realistic shape of daily temperatures one needs to introduce humidity in the picture.

The ultimate goal of this manuscript is to suggest a model that not only reproduces the three climatic effects above, but also predicts an extremely realistic evolution of average temperature and humidity. Only now accurate meteorological datasets of virtually any place in the world and for about 50 years are available. For this reason only now it is possible to compute a \emph{climatic} unfolding of temperatures and other meteorological parameters (e.g.\ humidity) for any region on our planet. 

To relate our work with the literature, we recall that in \cite{2012.AJ.Cowan.Voigt.Abbot} the authors discuss a very simple mathematical model to explain the phenomenon of lag of seasons. Elementary mathematics proves that the long-term solution to the equation
\begin{equation}\label{eq1}
\dot T(t)= (- \lambda T(t) + \mu) + (a \cos(\omega t) + b)
\end{equation}
is 
\begin{equation}\label{sol1}
\frac {\mu + b}\lambda  + \frac a{\sqrt{\lambda^2 + \omega^2}} \cos(\omega (t - \tau)) ,
\end{equation}
where $\tau = \varphi/\omega$ and $\varphi = \arg(\lambda + i \omega)$. Equation~\eqref{eq1} is an extremely basic model for the evolution of temperatures of a region $\mathcal R$ on the surface of a planet. In this model $T$ is the temperature of $\mathcal R$. The term $- \lambda T + \mu$ is a linearisation of the outgoing radiation from $\mathcal R$ while the forcing term $a \cos(\omega t) + b$ models the solar irradiance absorbed by $\mathcal R$. It clearly follows from \eqref{sol1} that the temperature $T$ has maxima and minima delayed with respect to the maxima and minima of solar irradiance, and the lag of these extreme temperatures is $\tau$.

The simple introduction of a forcing term containing two frequencies (daily rotation and annual revolution) is not rich enough to give realistic predictions of both lags (noons and seasons) \cite{2016.Ottolini}. To make such predictions one must increase the number of degrees of freedom. In fact, the first model whose solutions correctly predict both effects uses at least two different thermodynamic bodies with different thermic inertia, which correspond to a system with two degrees of freedom\cite{2017.DiBona}. A quantitative analysis of the two lags is performed in Appendix~\ref{appendice}. The asymmetry of temperatures is a further effect, and to be reproduced it requires the introduction of one more degree of freedom, that models the evolution of the absolute humidity of the air.
 

In the literature, models can be roughly divided into two categories: global circulation models (GCMs) \cite{2010.McGuffie.Henderson-Sellers} and energy balance models (EBMs) \cite{2012.AJ.Cowan.Voigt.Abbot}. In GCMs land, oceans, and atmosphere are discretised into cells, and flows and energy transfer among cells are integrated over time; in EBMs the evolution of temperature is computed through low-dimensional systems, and the investigation is typically local or mediated along a parallel. GCMs can predict climate more accurately, but they require great effort to acquire data, to set up the simulation, and need large computing capacities. EBMs are possibly less accurate but require much less computational resources. EBMs have often been used to investigate climate under hypothetical variations of orbital and environmental parameters \cite{2013.AJ.Vladilo.al, 2017.IJA.Silva.al}. Our work belongs to this second class of models. Our model could be used to investigate possible climate changes on Earth (e.g.\ greenhouse effect) and climate habitability of exoplanets in specific parts of their surface. In particular, at difference from classical one-dimensional EBMs \cite{2017.MNRAS.Silva.al}, our approach is applicable when the revolution period and the rotation period are in 1:1 resonance (tidal locking) or other low-order resonance.

The outline of the work is the following. In Section~\ref{geometria} we give some geometric definitions and we write explicitly the expression of solar rays inclination. In Section~\ref{fisica} we recall the general laws of heat exchange and evaporation, and we write the evolution equations for temperature and humidity of a planet's region. In Section~\ref{numerica} we numerically solve the equations for various regions on Earth, showing that our model well describes different types of \emph{climates} (according to the K\"oppen climate classification \cite{2018.EB.Arnfield}). In Section~\ref{conclusioni}, we discuss the results and we indicate possible improvements and applications of the model.

\section{The geometry of solar radiation}\label{geometria}

The motion and orientation in space of a region $\mathcal R$ on the surface of a planet is in good approximation due to the composition of the Keplerian \emph{revolution} of the planet around its star and the \emph{rotation} of the planet around its axis. The combination of such motions determines intensity and angle of the solar radiation responsible for the heating of the region. Disregarding all possible perturbations to this setting, the power of incoming solar radiation in $\mathcal R$ is hence completely determined by its exposition on the planet and the position of the planet in space. 

\subsection{Geometrical definitions}

In our model the planet is assumed to be spheric. Its center of mass, following Kepler's laws, revolves around the sun along an ellipse belonging to a plane called \emph{ecliptic plane}. The planet also rotates uniformly around an invariable axis which makes a fixed angle $\gamma$, called \textit{obliquity}, with respect to the normal of the ecliptic plane. The two points of the planet whose movement is not due to rotation are called \textit{North} and \textit{South poles}, and we agree that they are respectively at latitude $+90$ and $-90$ degrees (or $\pi/2$ and $-\pi/2$ radiants).  The \emph{tropics} are the two circles of points that have latitude $\pm \gamma$. We plan to describe the evolution of temperature in a certain region of the planet situated at a fixed latitude $\varphi$ and longitude $\psi$.

Astronomically speaking, significant instants are those in which the sun rays have local and global minimal (or maximal) distance from the zenit. Climatically speaking, significant instants are those in which the temperature has local and global maximum (or minimum). We hence give the definition of such events.

\begin{defi}
The \emph{solar solstices} are the two instant in which the sun is at the zenit in one of the two tropics. The \emph{thermal solstices} are the global extremes (maximum and minimum) of the temperature in a zone of the planet during the year.
\end{defi}

Let us observe that  at the solar solstice the following equivalent facts also take place:
 \begin{itemize}
\item the projection of the terrestrial axis on the ecliptic plane is along the planet-sun line, and the pole in the same hemisphere of the zone is exposed to the sun;
\item the sun is at the highest point when seen either from the North pole or from the South pole.
\end{itemize}

\begin{defi}
The \emph{solar noon} is the instant in which the sun is at the local maximal height with respect to the horizon. The \emph{thermic noon} is the moment in which the temperature is at a local maximum.
\end{defi}

As anticipated in the Introduction, two remarkable phenomena take place on Earth: the lag between solar and thermal solstices and the the lag between solar and thermal noons.
\begin{defi}
The \emph{lag of seasons} is the delay between the thermal and the solar solstice (summer and winter). The \emph{lag of noon} is the delay between the thermal and the solar noon.
\end{defi}

The astronomical special positions called \emph{aphelion} and \emph{perihelion} are unrelated to solstices. When the revolution of the planet around the star is not circular, we will have to keep into consideration the shift between summer/winter solstices and such aphelion/perihelion.

\subsection{Inclination of solar rays}\label{solar rays}

Let us consider a planet $P$ rotating around its sun $S$, and let $e_1,e_2,e_3$ be an orthonormal reference frame fixed with respect to the stars. The vector $e_1$ is parallel to the major semiaxes of the keplerian orbit of $P$ and is directed from $S$ to $P$ when $P$ is at the perihelion; the vector $e_3$ is normal to the ecliptic plane and is such that the rotation of $P$ around the sun is counterclockwise; the vector $e_2 = e_3\times e_1$ completes the frame and is parallel to the minor semiaxes.

Following the classical description of keplerian motions, and supposing that at time $t_0=0$ the planet $P$ is located at the perihelion, the position of $P$ with respect to the sun is given in polar coordinates by the formulas
\begin{equation}\label{kepler}
\rho(t) = \frac{a (1 - e^2)}{ 1 + e \cos(\vartheta(t))}, \qquad \dot \vartheta(t) = \frac {2 \pi}{Y \sqrt{1-e^2}^3}(1 + e \cos(\vartheta(t)))^2, \qquad \vartheta(0) = 0
\end{equation}

\par\noindent\begin{tabular}
{@{}p{7cm} p{7cm}}
where $e$ is the eccentricity, $a$ is the length of the major semi-axis of the orbit, and $Y$ is the period of revolution. We also suppose that the planet $P$ rotates with angular velocity $\Omega = 2\pi/D$ around an axis invariable in space ($D$ is the period of one rotation, also called \emph{sidereal day}). Such invariable axis can be determined by two angles, in fact the axis belongs to the cone that forms an angle $\gamma$ with $e_3$ and its projection on the $e_1,e_2$ plane forms an angle $\delta$ with the $e_1$-axis moving counterclockwise (see figure). This means that a convenient choice&
\begin{center}
\vskip -0.6cm
\hskip -2.5cm
\includegraphics[width=4.6cm,trim=1.8cm 3cm 4.5cm 1cm, clip=true]{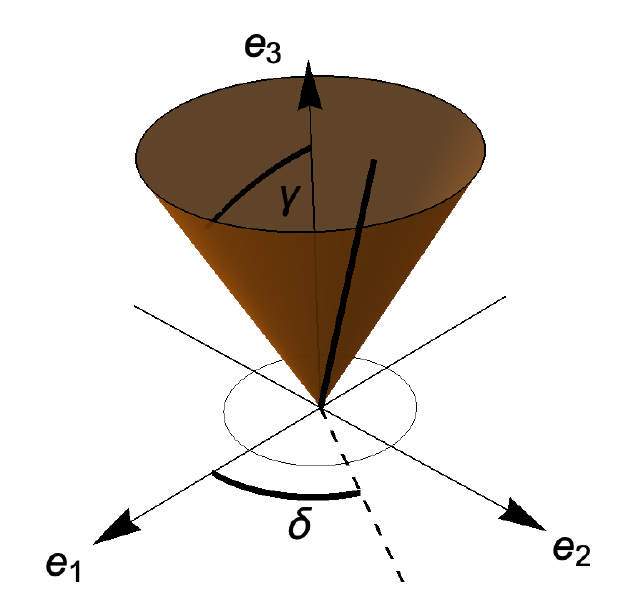}
\end{center}
\end{tabular}
 of reference frame $f_1,f_2,f_3$ attached to the rotating body with $f_3$ parallel to the axis of rotation is
%
\begin{equation*}
\begin{cases}
f_1(t) = e_1 (\cos\!\gamma \cos\!\delta \cos (\Omega t)-\sin\!\delta \sin (\Omega t)) + \\
\qquad \qquad + e_2 (\cos\!\gamma \sin\!\delta \cos (\Omega t)+\cos\!\delta \sin (\Omega t))-e_3 \sin\!\gamma \cos (\Omega t)\\[5pt]
f_2(t) = e_1 (\sin\!\delta (-\cos (\Omega t))-\cos\!\gamma \cos\!\delta \sin (\Omega t)) + \\
\qquad \qquad +e_2 (\cos\!\delta \cos (\Omega t)-\cos\!\gamma \sin\!\delta \sin (\Omega t))+e_3 \sin\!\gamma \sin (\Omega t)\\[5pt]
f_3(t) = e_2 \sin\!\gamma \sin\!\delta+e_1 \sin\!\gamma \cos\!\delta+e_3 \cos\!\gamma.
\end{cases}
\end{equation*}
The versor connecting the planet to the sun is 
\[
d(t) = - \cos(\vartheta(t)) e_1 - \sin(\vartheta(t)) e_2.
\]
Since the region $\mathcal R$ at latitude $\varphi$ and longitude $\psi$ has normal to the surface 
\begin{equation*}
n(t) = \cos(\psi) \cos\!\varphi f_1(t) + \sin(\psi) \cos\!\varphi f_2(t) + \sin\!\varphi f_3(t),
\end{equation*}
it follows that
\begin{multline}\label{trigonometria e raggi solari}
n(t) \cdot d(t) = \cos\!\varphi \sin (\delta - \vartheta(t) ) \sin (\Omega t + \psi ) + \\
- \cos (\delta -\vartheta(t)) \big(\sin\!\gamma \sin\!\varphi+\cos\!\gamma \cos\!\varphi \cos (\Omega t  + \psi)\big).
\end{multline}
This scalar product will be used in the following section, when writing the solar irradiance. In Table~\ref{table:astronomical parameters} the values of all parameters used in this discussion are indicated when the planet is the Earth.
 


\begin{table}[h]
\begin{center}
\begin{tabular}{|c|c|c|}
\hline
$e$ & $0.0167$ & Eccentricity of Earth's orbit\\
$a$ & $1.496\times 10^{11} \si{\meter}$ & Average earth-sun distance\\
$\gamma$ & $23.437\si{\degree}$ & Earth's mean obliquity\\
$\delta$ & $-12.8219\si{\degree}$ & Angle between solstices and perihelion/aphelion\\
$D$ & $8.616409 \times 10^4 \si{\second}$ & Period of rotation of the Earth\\
$Y$ & $3.15569 \times 10^7 \si{\second}$ & Period of revolution of the Earth\\
\hline
\end{tabular}
\end{center}
\caption{Fundamental astronomical parameters for planet Earth.}
\label{table:astronomical parameters}
\end{table}

\section{The physics of heat transfer}\label{fisica}

The temperature of a region $\mathcal R$ on a planet is the result of a balance between the incoming radiation from the sun, the outgoing radiating energy, latent heats and heat exchanges within the system. We model the heat dynamics of such limited region $\mathcal{R}$ located at a certain latitude $\varphi$ and a certain longitude $\psi$. We disregard spatial diffusion and hence we use ordinary differential equations in which time is the independent variable. This is not a reasonable assumption when dealing with meteorology, but climatology deals with average evolution of the temperatures and the influence of neighbouring regions should average out. It follows that, to model experimental mean data, it is reasonable to suppose that the region $\mathcal R$ is physically isolated from the rest of the planet.

We restrict our study to the lowest part of the atmosphere and to the superficial layer of the planet's surface. As we said in the Introduction, in order to reproduce lags, daily patterns, and more generally local climates, we consider three different homogeneous thermodynamic bodies, that in the case of Earth are air (which temperature we measure), land, and sea. To keep the model simple, we consider a unique mixed layer of air \cite{2010.Pierrehumbert} for the lower atmosphere.  We refer to the air layer using the index 0, to the land using the index 1, and to the ocean using the index 2.

This model represents the energy balance of the region $\mathcal R$ of the Earth's surface which extension is reasonably of the order of $100\si{\kilo\meter^2}$. The real value of this surface plays no role in our investigation. In fact all quantities will be expressed ``per unit surface", and the units will always be divided per $\si{\meter}^{-2}$.

\subsection{Solar irradiance}\label{solar radiation}

Approximating the sun to a black body, the solar irradiance flowing through a unit area perpendicular to the rays at distance $\rho$ from the sun is given by the Stefan-Boltzmann law
\begin{equation}\label{Stefan-Boltzmann}
I = \sigma T_s^4 \frac{R_s^2}{\rho^2}.
\end{equation}
Here $\sigma$ is the Stephan-Boltzmann constant ($\si{\joule \second^{-1}\kelvin^{-4}\meter^{-2}}$), $R_s$ is the radius of the sun, $T_s$ is the temperature of the sun (see Table~\ref{table:solar irradiance parameters}). In order to have the effective power received by a unit region $\mathcal R$ on the planet, we must multiply \eqref{Stefan-Boltzmann} by the scalar product \eqref{trigonometria e raggi solari}. Considering the fact that during the night the contribution of the solar radiation is zero, the solar irradiance on $\mathcal R$ is
\begin{equation}\label{power}
W(t) = \max\left\{\sigma T_s^4 \frac{R_s^2}{\rho(t)^2}\; n(t) \cdot d(t),\: 0 \right\}.
\end{equation}

Observe that this quantity is expressed in $\si{\joule \second^{-1} \meter^{-2}}$ and is the power of solar irradiance per unit area. When a light ray hits a body, its energy can be absorbed, transmitted or reflected. These three phenomena can be modelled introducing three parameters: \emph{absorbance} $\alpha$, \emph{transmittance} $\tau$, and \emph{reflectance} $r$ such that $\alpha + \tau + r = 1$. We mention here that in the literature the fraction of reflected radiation is commonly called \emph{albedo}.

The solar rays cross the whole atmosphere, which absorbs a part of them. When the rays reach the surface, a part of them is absorbed by the superficial layer, another part is transmitted to a deeper underlying layer and a last part is reflected back to the atmosphere. Again, a part of this reflected radiation is absorbed, reflected or transmitted by the atmosphere. The layer of land and sea absorb all the incoming radiation but in a very different way; for this reason we must keep in mind that each region $\mathcal R$ is partly land and partly water. For this reason we introduce the main climatic parameter: a number \(p \in [0,1]\) which represents the fraction of land and is referred to as \emph{solid fraction parameter}. Its complementary parameter $q = 1 - p$ is the fraction of ocean.

The quantity of solar radiation absorbed by the three layers follows the laws
\begin{equation}\label{SR}
\begin{cases}
\dfrac{dQ_0^{SR}}{dt} = \alpha_0(1 + p \tau_0 r_1 + q \tau_0 r_2)  W(t)\\[8pt]
\dfrac{dQ_1^{SR}}{dt} = p \tau_0 \alpha_1 W(t)\\[8pt]
\dfrac{dQ_2^{SR}}{dt} = q\tau_0 \alpha_2 W(t).
\end{cases}
\end{equation}
The superscript $SR$ indicates that the contribution comes from Solar Radiation. The quantities $Q_i$ are expressed in $\si{\joule \, \meter^{-2}}$ and represent the heat quantity of the three thermodynamic bodies per unit area. The true amount of energy stored in such bodies can be obtained multiplying by the surface taken into consideration.

\begin{table}[h]
\begin{center}
\begin{tabular}{|c|c|c|}
\hline
Parameter & Value & Description\\
\hline
$\sigma$ & $5.670\times10^{-8} \, \si{\joule\second^{-1}\kelvin^{-4}\meter^{-2}}$  & Stefan-Boltzmann constant\\
$R_S$ & $6.955\times 10^8 \, \si{\meter}$ & Solar radius\\
$T_S$ & $5778 \, \si{\kelvin}$ & Sun superficial temperature\\
\hline
\end{tabular}
\end{center}
\caption{Fundamental physical parameters for the solar irradiance.}
\label{table:solar irradiance parameters}
\end{table}

The parameters $\tau_i, \alpha_i, r_i = 1 - \alpha_i - \tau_i,p, q = 1-p$ are considered constants. We are aware that they actually are slightly variable, depending on the zenith distance of the sun, the atmosphere composition, the superficial temperature, and other factors. We will use their average value in the numeric integration.
In our simulations, we have chosen $r_1 = 0.2$ for the reflectance of the land, which is a good approximation for Earth continents \cite{1997.I.Williams.Kasting}. For other types of surface we can consider values of $r_1$ in the range $[0.1,\,0.4]$ \cite{2015.AJ.Vladilo.al}. The lowest values are appropriate for basaltic rocks or conifer forests, Sahara's desert has $r_1 \simeq 0.4$ \cite{2008.T.Muller.al}, while grasslands have $r_1 \simeq 0.2$ \cite{2010.Pierrehumbert}. With respect to the ice, it has been documented a difference between ices over lands and over oceans \cite{2010.Pierrehumbert, 1969.Kondratyev}. Therefore, following \cite{2015.AJ.Vladilo.al}, we adopt $r_1 = 0.85$ and $r_2 = 0.62$ for ices over lands and ices over oceans respectively. We also suppose that all of the solar radiation not reflected by the surface is absorbed, giving $\alpha_1 = 1-r_1$, and $\alpha_2 = 1-r_2$. For the atmosphere the absorbance of solar radiation $\alpha_{0}$ is slightly variable \cite{2009.AMS.Tremberth.Fasullo.Kiehl}, we assign to it the average values $0.25$. The transmittances $\tau_i$ are given by the relation $\tau_i  = 1 - \alpha_i - r_i$.
In Table~\ref{table:solar radiation parameters} the values of all relevant parameters are listed.


%

\begin{table}[h]
	\begin{center}
		\begin{tabular}{|c|c|c|c|c|}
			\hline
			 & Reflectance $r$ & Absorbance $\alpha$ & Transmittance $\tau$ \\
			\hline
			Atmosphere (solar light)& 0.23 & $0.25$ & 0.52\\
			Soil & 0.2 & 0.8 & 0\\
			Desert & 0.4 & 0.6& 0 \\
			Ocean & 0.15 & 0.85& 0 \\
			Ice over land & 0.85 & 0.15& 0 \\
			Ice over oceans & 0.62 & 0.38& 0 \\
			\hline
		\end{tabular}
	\end{center}
	\caption{Reflectance $r$, absorbance $\alpha$ and transmittance $\tau$ of solar radiation for various thermodynamic bodies on Earth.}
	\label{table:solar radiation parameters}
\end{table}

\subsection{Thermal radiation}\label{thermal radiation} 
All hot objects radiate with a Stefan-Boltzmann law. Unlike the sun, warm objects cannot be assumed to be black bodies and hence the power of emitted energy is $\varepsilon \sigma T^4$, where $\varepsilon$ is the \emph{emissivity} of the body, a number in $[0, 1]$ which depends on chemical and physical properties of the hot body. In this model the atmosphere will be assumed to radiate in two directions, down towards the earth with emissivity $\varepsilon_0^d$, and up towards outer space with emissivity $\varepsilon_0^u$. We also assume that $\varepsilon_0^d > \varepsilon_0^u$ because of lower density and temperature of the upper part of the atmosphere, and that all downward infrared radiation is absorbed by soil and water. We choose $\varepsilon_0^d = 0.8$ for the radiation to the earth surface and $\varepsilon_0^u = 0.45$ for the radiation to outer space. 

The correct energy balance at our temperatures must include a parameter $\alpha_{0}^T$ to model the absorbance by the atmosphere of the radiation, called \emph{thermal radiation}, emitted from Earth \cite{2009.AMS.Tremberth.Fasullo.Kiehl}. Unlike solar radiation, the spectrum of thermal radiation is mainly infrared, and $\alpha_{0}^{T}$ is much higher than $\alpha_0$. The value assigned to $\alpha_0^T$ is connected to the modelling of the greenhouse effect and it belongs to the interval $[0.8,0.95]$.

Summarizing,  the power of energy transferred through thermal radiation between the thermodynamic bodies in $\mathcal R$ is
\begin{equation}\label{TR}
\begin{cases}
\dfrac{dQ_0^{TR}}{dt}= \sigma (p\alpha_{0}^{T}\varepsilon_1 T_1^4 +q\alpha_0^T\varepsilon_2 T_2^4 - (\varepsilon_0^d + \varepsilon_0^u) T_0^4)\\[8pt]
\dfrac{dQ_1^{TR}}{dt}= p \sigma (\varepsilon_0^d T_0^4 - \varepsilon_1 T_1^4) \\[8pt]
\dfrac{dQ_2^{TR}}{dt}= q \sigma (\varepsilon_0^d T_0^4 - \varepsilon_2 T_2^4).
\end{cases}
\end{equation}
The superscript TR stands for Thermal Radiation. 
For the thermal radiation, we consider these values of emissivity $\varepsilon_1^{soil}= 0.94$ for soil, $\varepsilon_1^{sand} = 0.75$ for deserts, $\varepsilon_2 = 0.96$ for oceans, and $\varepsilon_1^{ice} = \varepsilon_2^{ice} = 0.85$ for ices over land and over ocean \cite{2009.ASHRAE.AAVV}. We suppose that the atmosphere absorbs most of the radiation emitted by the surface. All values are summarised in Table~\ref{table:thermal radiation parameters}.

\begin{table}[h]
	\begin{center}
		\begin{tabular}{|c|c|c|}
			\hline
			Parameter & Value & Description\\
			\hline
			$\alpha_0^T$  & [0.8, 0.95] & Atmosphere absorbance (infrared light)\\
			$\varepsilon_0^d$ & 0.8 & Atmosphere emissivity downwards\\
			$\varepsilon_0^u$ & 0.45 & Atmosphere emissivity upwards\\
			\hline
			$\varepsilon_1^{soil}$ & 0.94 & Soil emissivity\\
			$\varepsilon_1^{sand}$ & 0.75 & Sand emissivity\\
			$\varepsilon_1^{ice}$ & 0.85 & Ice over land emissivity \\
			\hline
			$\varepsilon_2^{water}$ &  0.96 & Water emissivity\\
			$\varepsilon_2^{ice}$ & 0.85 & Ice over oceans emissivity\\
			\hline
		\end{tabular}
	\end{center}
	\caption{Atmospheric absorbance $\alpha_0^T$ to thermal radiation and emissivity $\varepsilon$ for various thermodynamic bodies on Earth.}
	\label{table:thermal radiation parameters}
\end{table}

\subsection{Conduction and convection}\label{CCE}

According to Fourier's law, the rate at which two warm bodies exchange heat is proportional to the negative gradient of the temperature and to the area through which the heat flows. A similar law exists for convection, and is called Newton's law of cooling. Altogether, if $T_1$ and $T_2$ are the temperatures of the two thermodynamic bodies, the heat flow $Q$ due to conduction and convection between them follows the law
\begin{equation}
\frac{dQ}{dt} = h(T_2-T_1),
\end{equation}
where $h$ is the cumulative heat transfer coefficient. In our model, the contributions of heat exchange due to conduction and convection are
\begin{equation}\label{C}
\begin{cases}
\dfrac{dQ_0^C}{dt}=p h_{01}(T_1(t)-T_0(t))+qh_{02}(T_2(t)-T_0(t))\\[8pt]
\dfrac{dQ_1^C}{dt}=-p h_{01}(T_1(t)-T_0(t))\\[8pt]
\dfrac{dQ_2^C}{dt}=-q h_{02}(T_2(t)-T_0(t)),
\end{cases}
\end{equation}
where $h_{ij}$ is the heat transfer coefficient among the two components labelled $i$ and $j$. In Table~\ref{table:heat transfer coefficients} the range for such coefficients are reported.

\begin{table}[h]
\begin{center}
\begin{tabular}{|c|c|c|}
\hline
Parameter & Value & Description\\
\hline
$h_{01}$ & $[5, 40] \, \si{\joule\second^{-1}\meter^{-2}}$ & land-air heat transfer coefficient\\
$h_{02}$ & $[5, 40] \, \si{\joule\second^{-1}\meter^{-2}}$ & water-air heat transfer coefficient\\
\hline
\end{tabular}
\end{center}
\caption{Heat transfer coefficients between air and land and air and water.}
\label{table:heat transfer coefficients}
\end{table}


\subsection{Geothermal heat.}
In our model we take into consideration geothermal energy, that is heat coming from the mantle. There is a well defined region separating the mantle from the planet's crust, called \textit{Mohorovi\v{c}i\'c discontinuity }or \textit{Moho}. Since the temperature of the mantle is much higher than the temperatures on the surface, we can assume that the geothermal heat flow is constant and we write
\begin{equation}\label{M}
\dfrac{dQ_1^M}{dt}= p \, \eta_1, \qquad \dfrac{dQ_2^{M}}{dt}= q \, \eta_2, \qquad \dfrac{dQ_0^M}{dt}=0.
\end{equation}

The parameter $\eta_i$ is the power of energy conducted from the mantle to the body $i$ per unit area. Using experimental data from 20201 sites covering 62\% of the Earth's surface, Pollack et al. in \cite{1993.RG.Pollack.Hurter.Johnson} have obtained the values shown in Table~\ref{table:geothermal power}.

\begin{table}[h]
\begin{center}
\begin{tabular}{|c|c|c|}
\hline
Parameter & Value & Description\\
\hline
$\eta_1$ & $0.345 \, \si{\joule\second^{-1}\meter^{-2}}$ & continental geothermal power\\
$\eta_2$ & $0.802 \, \si{\joule\second^{-1}\meter^{-2}}$ & oceanic geothermal power\\
\hline
\end{tabular}
\end{center}
\caption{Geothermal powers.}
\label{table:geothermal power}
\end{table}

The contribution of geothermal heat is between two and three order of magnitude lower than the contribution given by solar radiation. Its effect is hence feeble on Earth.

\subsection{Evaporation.}

The evaporation is a phenomenon that effects the absolute humidity of the air, and it depends on the wind speed and on the difference between the absolute humidity (the amount of kilograms of water vapour that a kilogram of dry air contains) and the saturation humidity of the air (the amount of kilograms of water vapour that a kilogram of dry air can contain at saturation). Saturated air cannot absorb water vapour, dry air does absorb vapour faster. We will assume that evaporation from land and sea is given by the law $\mu_i  (U_s(T_0) - U(t))$, with $\mu_i $ the rate of evaporation from the land and the sea expressed as a frequency ($\si{\second^{-1}}$). We assume the parameter $\mu_2$ to be variable depending on the wind speed. We consider the following approximation \cite{2004.Engineering_ToolBox, 2016.JAMTP.Orvos.Szabo.Poos}:
\begin{equation}\label{eq:approximation rate evaporation}
\mu_2 = \frac{25+19 \overline{v}_i}{3600}.
\end{equation}
The parameter $\mu_1$ is very variable, from very low values in the desert to very high values in tropical forests.

Measurements on Earth's surface indicate that averaged wind speed is very different from one place to another, varying in $[0.17,\,20]\,\si{\meter/\second}$ \cite{2015.Grassi.Veronesi.Schenkel.et.al} at 10 $\si{\meter}$ above the surface. When not directly accessible, we will use a reference value  $\overline{v}_i = 4\,\si{\meter/\second}$, which gives $\mu_2 = 5.7 \times 10^{-5} \, \si{\second^{-1}}$.

Another factor that subtracts water vapour from the atmosphere is rain. The physical process that causes rain is condensation when the moist air rises to higher and colder strata of the atmosphere. We model this effect assuming a rate of rainfall proportional to the absolute humidity of the low atmosphere, and we call the coefficient of proportionality $\nu$, espressed in $\si{\second^{-1}}$. The values of $\nu$ can be computed knowing average rain precipitation in a year $\Pi$ (in meter of rain per square-meter), average humidity of the air $\overline U$ in kilograms of water vapour per kilogram of dry air. The parameter $\nu$ can be obtained using the formula
\[
\nu \overline U Y \rho_0 \ell_0= \rho_2 \Pi.
\]
We indicate with $\rho_0$ the density of air, with $\rho_2$ the density of water, and with $\ell_0$ the depth of the atmospheric layer. It turns out that reasonable values for $\nu$ are of the order of $10^{-5}$.

\begin{figure}[h]\label{fig:saturation humidity}
\begin{center}
\includegraphics[width = 8cm]{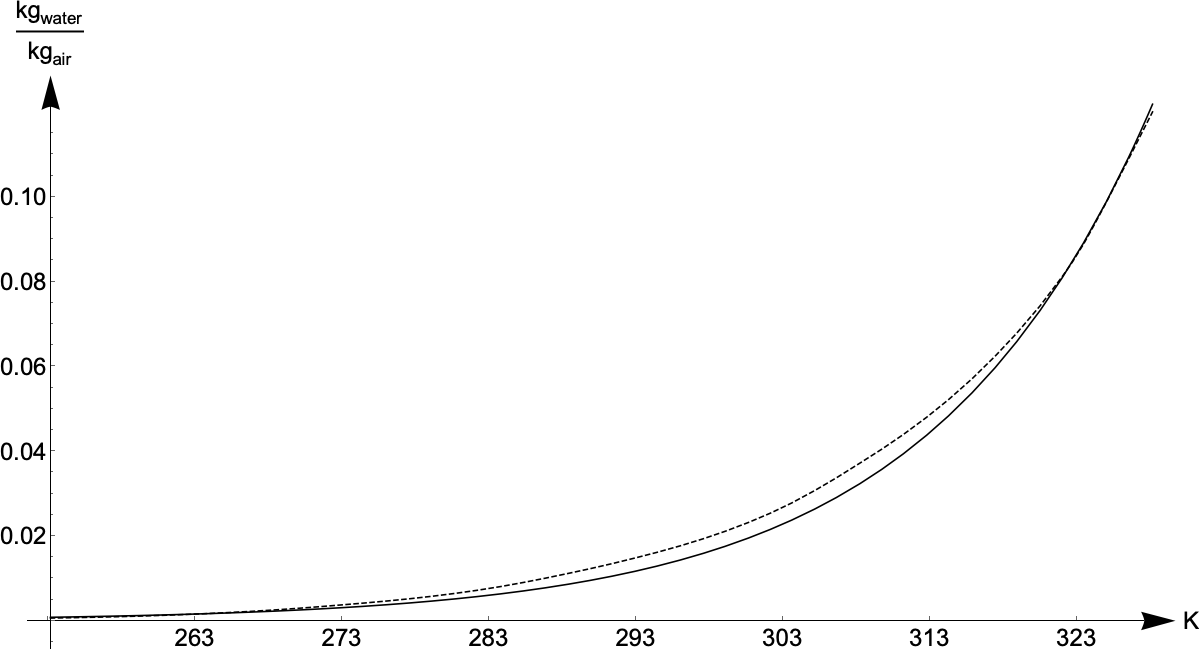}
\end{center}
\caption{Plot of the saturation humidity function $U_s(T)$, $T$ is expressed in Kelvin. The dotted line represents the empirically measured values, the continuous line is the exponential approximation \eqref{eq:us}.}
\end{figure}

Summarizing, the law that regulates the evolution of humidity in the air is
\begin{equation}\label{Evaporation}
\begin{split}
\dfrac{dU}{dt}(t) =  p \mu_1  (U_s(T_0(t)) - U(t)) + q \mu_2 (U_s(T_0(t)) - U(t)) - \nu U = \\
= (p \mu_1 + q \mu_2)  (U_s(T_0(t)) - U) - \nu U
\end{split}
\end{equation}
where
\begin{equation}\label{eq:us}
U_s(T) = e^{0.0666 T-23.96}
 \end{equation}
is the humidity of saturation, a function of the air temperature $T$ whose values are the maximal amount of $\si{\kilo\gram}$ of water vapour that a $\si{\kilo\gram}$ of dry air can contain. This function has been obtained fitting well known values, its graph is represented in Figure~\ref{fig:saturation humidity}.

\begin{table}[t]
\begin{center}
\begin{tabular}{|c|c|c|}
\hline
Parameter & Value & Description\\
\hline
$\mu_2$ & $5.7 \times 10^{-5} \, \si{\second^{-1}}$ & Evaporation rate from water\\
$\mu_1$ & $[10^{-6} ,10^{-4}] \, \si{\second^{-1}}$ & Evaporation rates from land\\
$\nu$ & $[1,5]\times 10^{-5} \, \si{\second^{-1}}$ & Rainfall rate\\
$\lambda$ & $2.26\times 10^6 \, \si{\joule \kilogram^{-1}}$ & Latent heat of evaporation and condensation\\
\hline
\end{tabular}
\end{center}
\caption{Evaporation rate, rainfall rate, and latent heat.}
\label{table:evaporation}
\end{table}

\subsection{Latent heat of evaporation and condensation.}

As we have seen in last section, humidity plays a crucial role in the thermodynamics of the system under investigation. In fact, given a certain temperature and atmospheric composition, evaporation and condensation of water take place, depending on the difference between the absolute humidity and the saturation humidity. As we know, for each phase transition there is a latent heat, that is heat used for phase transition. During the process of evaporation part of the solar energy is used to change from liquid to vapour phase. That energy is not used to increase the temperature of the thermodynamic body. Therefore, if the mass of water undergoing evaporation per unit time and area is given by $\rho_0\ell_0 dU/dt >0$, the related latent heat of evaporation is given by the formula
\begin{equation}\label{eq:evaporation latent heat}
	\dfrac{dQ_1^{LE}}{dt} =  - p\lambda \max\left(\rho_0\ell_0 \frac{dU}{dt},\,0\right), \qquad 
	\dfrac{dQ_2^{LE}}{dt} =  - q\lambda \max\left(\rho_0\ell_0 \frac{dU}{dt},\,0\right),
\end{equation}
where $\lambda$ is the specific latent heat for evaporation of water \cite{2011.Singh.Singh.Haritashya}. If $dU/dt<0$, the opposite process, called condensation, takes place. During this process, heat is released to the environment, with the same law as that for evaporation. In our system, the latent heat of condensation is released directly to the atmosphere, with the law
\begin{equation}\label{eq:condensation latent heat}
	\dfrac{dQ_0^{LC}}{dt} =  - \lambda \min(\rho_0\ell_0 \frac{dU}{dt},\,0).
\end{equation}
Averaging during the year, it is known that heat exchanged through these processes amounts at about 25\% of the solar irradiance \cite{2013.Wild.Folini.Schar.et.al}. To compare the magnitude of this process, heat transfer through convection amounts to about $5\%$ of the solar irradiance, and the energy absorbed directly by the atmosphere is between $18\%$ and $25\%$ of solar irradiance.

\subsection{Thermal inertia.}
Under the effect of heat transfers, the rate at which the temperature of a thermodynamic body change depends on its thermal capacity. In our case
\begin{equation}\label{Thermal capacity}
\frac{dQ_0}{dt}= C_0(U) \frac{dT_0}{dt}, \qquad \frac{dQ_1}{dt}= p C_1 \frac{dT_1}{dt}, \qquad \frac{dQ_2}{dt}= q C_2 \frac{dT_2}{dt}.
\end{equation}
The parameters $C_i$ are the thermal capacities per unit surface ($\si{\joule \kelvin^{-1}\meter^{-2}}$). For a body $i$ with density $\rho_i$ ($\si{\kilogram\per\meter^3}$), specific heat capacity $c_i$ ($ \si{\joule\per\kelvin\per\kilogram}$), and depth $\ell_i$, the thermal capacity per unit surface is $C_i = \rho_i\,c_i\,\ell_i$. We will assume the thermal capacities constant for all thermodynamic bodies except for the air. This is justified by the fact that the thermal capacity of the air depends on its content of water vapour. Recalling that $U$ is the absolute humidity of the air, measured in $\si{\kilogram}$ of water per $\si{\kilogram}$ of air, we will assume that the heat capacity of the air is
\[
C_0(U) =  c_0^d \rho^d_0 \ell_0 + c_0^v \rho_0^d \ell_0 U = C_0^d + C_0^v U,
\]
where $c_0^d$ is the specific heat capacity of dry air, and $c_0^v$ is the specific heat capacity of water vapour, $\rho_0^d$ is the density of dry air, and $\ell_0$ is the effective depth of the layer of air. The new independent variable $U$ here introduced will in turn depend, via a differential equation, from the temperature of the air. The specific heat of dry air is $711.28\,\si{\joule\per\kilogram\per\kelvin}$, the specific heat of water vapour is $2050.16\, \si{\joule\per\kilogram\per\kelvin}$. In our model we consider a layer of lower atmosphere $\ell_0 = 400\, \si{\meter}$.

Following \cite{2010.Pierrehumbert, 1997.I.Williams.Kasting}, we choose $\ell_1 \in [0.3,\, 0.5]\, \si{\meter}$ for soil, $\ell_2 \in [40,\, 60]\, \si{\meter}$ for oceans. The thermal characteristics of land and water differ from region to region. For this reason in different cases we use different heat capacities. For details on such values see \cite{2015.AJ.Vladilo.al, 2015.EP.Iniesta.al, 1994.GF.Cuffey.Patterson, 2015.EP.Diago}. Using the arguments above one obtains the values shown in Table~\ref{table:heat capacities}.


\begin{table}[t]
\begin{center}
\begin{tabular}{|c|c|c|}
\hline
Parameter & Value  & Description\\
\hline
$C_0^d$ & $3.5 \times 10^5 \, \si{\joule\kelvin^{-1}\meter^{-2}}$ & dry air thermal capacity\\
$C_0^v$ & $1 \times 10^6 \, \si{\joule\kelvin^{-1}\meter^{-2}}$ & water vapor thermal capacity\\
\hline
$C_1^{soil}$ & $1.0 \times 10^6 \, \si{\joule\kelvin^{-1}\meter^{-2}}$ & soil thermal capacity\\
$C_1^{ice}$ & $1.0\times 10^6 \, \si{\joule\kelvin^{-1}\meter^{-2}}$ & ice thermal capacity \\
$C_1^{forest}$ & $1.7 \times 10^6 \, \si{\joule\kelvin^{-1}\meter^{-2}}$ & forest thermal capacity\\
$C_1^{sand}$ & $3.2 \times 10^6 \, \si{\joule\kelvin^{-1}\meter^{-2}}$ & sand thermal capacity \\
\hline
$C_2$ & $[1.7,\, 2.5] \times 10^8 \, \si{\joule\kelvin^{-1}\meter^{-2}}$ & ocean thermal capacity\\
\hline
\end{tabular}
\end{center}
\caption{Heat capacities.}
\label{table:heat capacities}
\end{table}
 
\subsection{Final system.}

Summarizing equations~\eqref{kepler}, \eqref{trigonometria e raggi solari}, \eqref{power}, \eqref{SR},  \eqref{TR}, \eqref{C}, \eqref{M}, \eqref{Thermal capacity}, \eqref{Evaporation}, \eqref{eq:evaporation latent heat}, \eqref{eq:condensation latent heat}, with some minimal algebra, the dynamical system that models the temperature evolution of $\mathcal{R}$ is modelled by the evolution of the 4 independent variables $T_0,T_1,T_2,U$, and the variable $\vartheta$, whose evolution is fixed for planet Earth,
\begin{equation*}\label{2RCM}
\begin{cases}
(C_0^d + U C_0^v)\dfrac{dT_0}{dt} =\alpha_0(1 + p \tau_0 r_1 + q \tau_0 r_2)   W +  \sigma \alpha_0^T (p \varepsilon_1 T_1^4 + q \varepsilon_2 T_2^4)  +
\\[5pt]
\hskip 0.7cm - \sigma (\varepsilon_0^d + \varepsilon_0^u) T_0^4) + p h_{01}(T_1 - T_0)+qh_{02}(T_2 - T_0) - \lambda^E \min(\rho_0\ell_0 \frac{dU}{dt},\,0).
\\[6pt]
C_1\dfrac{dT_1}{dt} =  \tau_0 \alpha_1 W + \sigma (\varepsilon_0^d  T_0^4 - \varepsilon_1 T_1^4) - h_{01}(T_1 - T_0) + \eta_1 - \lambda^E \max(\rho_0\ell_0 \frac{dU}{dt},\,0)
\\[12pt]
C_2\dfrac{dT_2}{dt} = \tau_0 \alpha_2 W + \sigma (\varepsilon_0^d T_0^4 - \varepsilon_2 T_2^4)  - h_{02}(T_2 - T_0) + \eta_2 - \lambda^E \max(\rho_0\ell_0 \frac{dU}{dt},\,0)
\\[12pt]
\dfrac{dU}{dt} = (p \mu_1 + q \mu_2) U_s\circ T_0 - (p \mu_1 + q \mu_2 + \nu) U \\[12pt]
\dfrac{d\vartheta}{dt} = \dfrac {2 \pi}{Y \sqrt{1-e^2}^3}(1 + e \cos\!\vartheta)^2.
\end{cases}
\end{equation*}



\section{Numerical analysis}\label{numerica}

In this section we justify the final choice of the parameters depending on the choice of region $\mathcal R$ on Earth, and we run the simulation for various types of climates. We then compare the numerical results with the real average temperatures. The numerics, the acquisition of real temperatures, and their manipulation have been done using the software Mathematica Wolfram Research Inc. In particular \texttt{WeatherData[]} allowed us to acquire the dataseries of temperatures and humidity with respect to coordinated universal time (UTC) from a variety of weather stations in regions with different climates. We averaged the temperatures at any given time of the year over a period of 47 years (from 1973 to 2019).

We consider 5 regions: Hilo--Hawaii, Kufra--Lybia, Catania--Italy, Lincoln--USA,Vostok--Antarctica. Each region belongs to one of the K\"oppen climate zones \cite{2018.EB.Arnfield}: Tropical (A), Arid (B), Temperate (C), Continental (D), and Polar (E). In the following subsections we choose the local parameters and we superimpose mean temperature and humidity in the chosen region with temperature and humidity obtained with our model. In the yearly plots we indicate with dashed lines the solstices and equinoxes; in the daily plots we indicate midday and midnights.

%

\subsection{Tropical climate: Hilo, Hawaii}
\begin{figure}[h]\begin{center}
\includegraphics[width = 0.5 \textwidth]{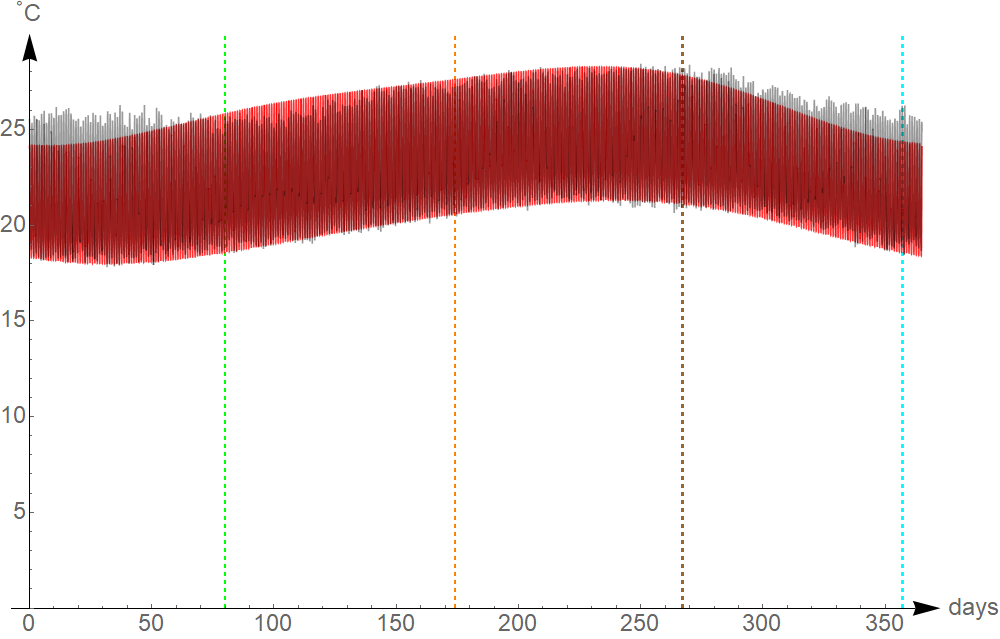}\includegraphics[width = 0.5 \textwidth]{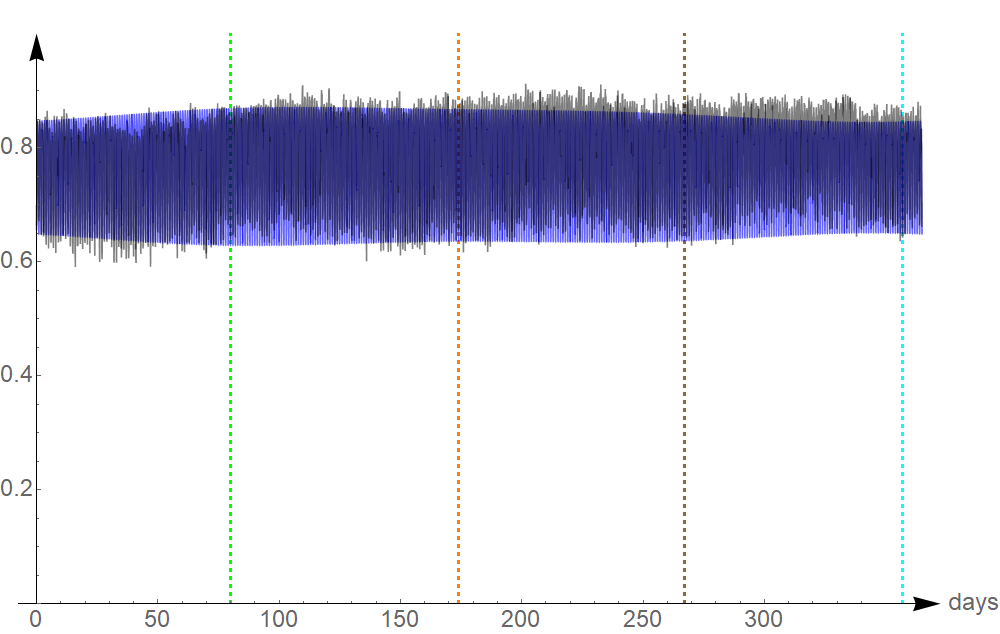}
\includegraphics[width = 0.25 \textwidth]{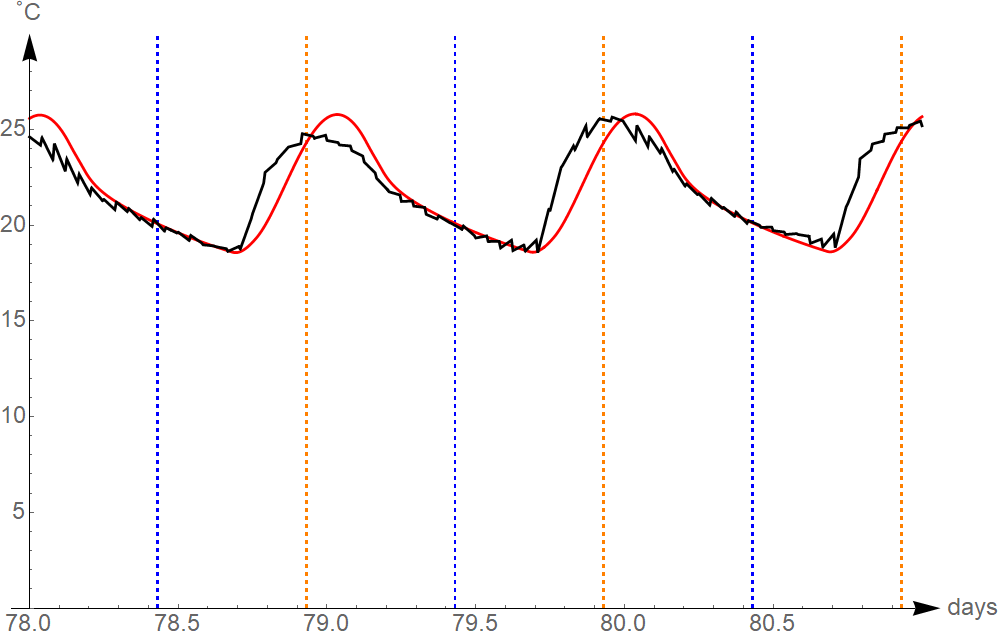}\includegraphics[width = 0.25 \textwidth]{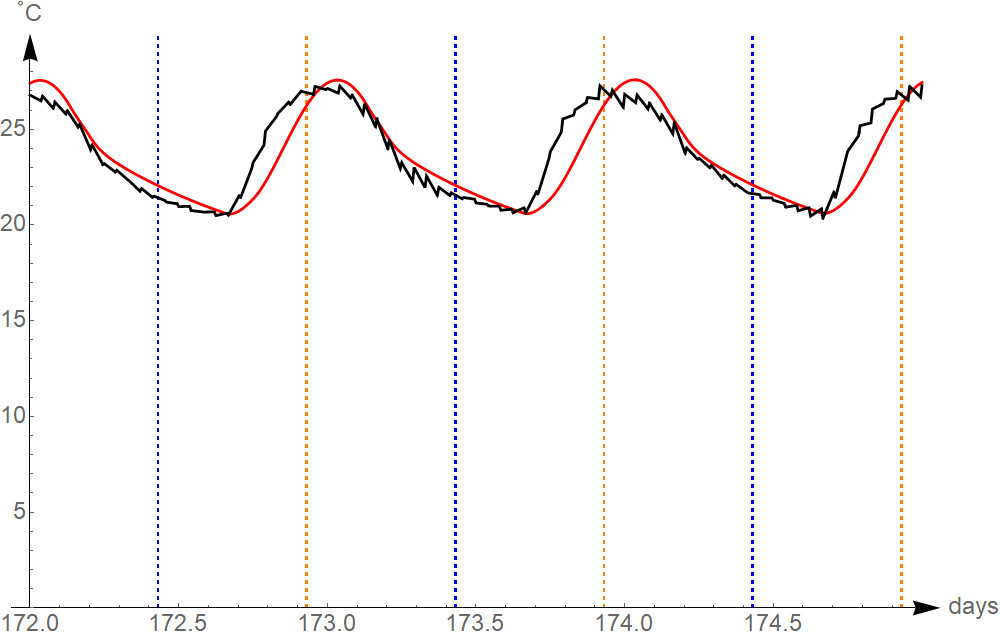}\includegraphics[width = 0.25 \textwidth]{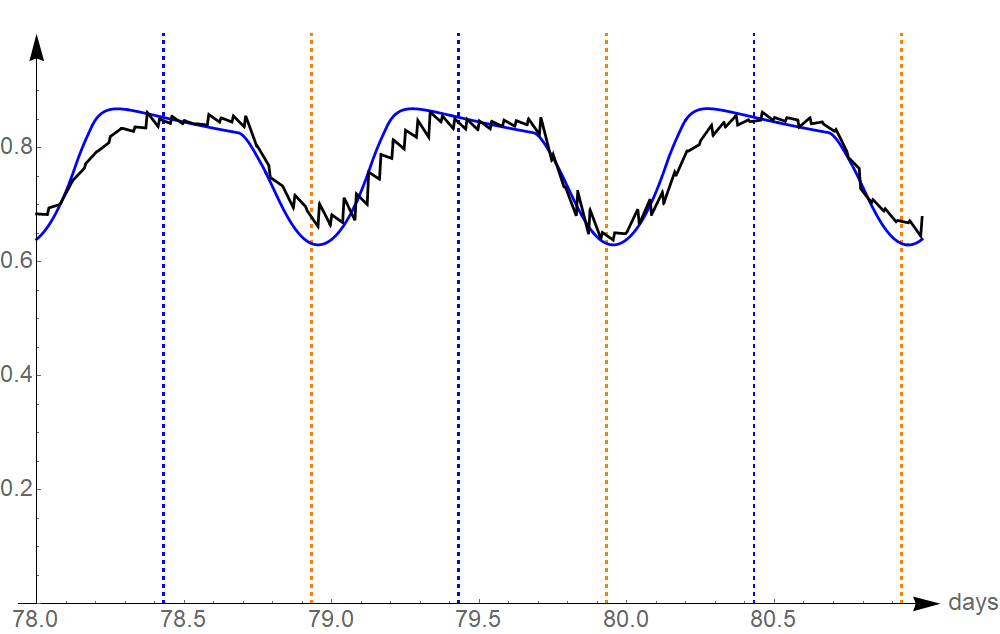}\includegraphics[width = 0.25 \textwidth]{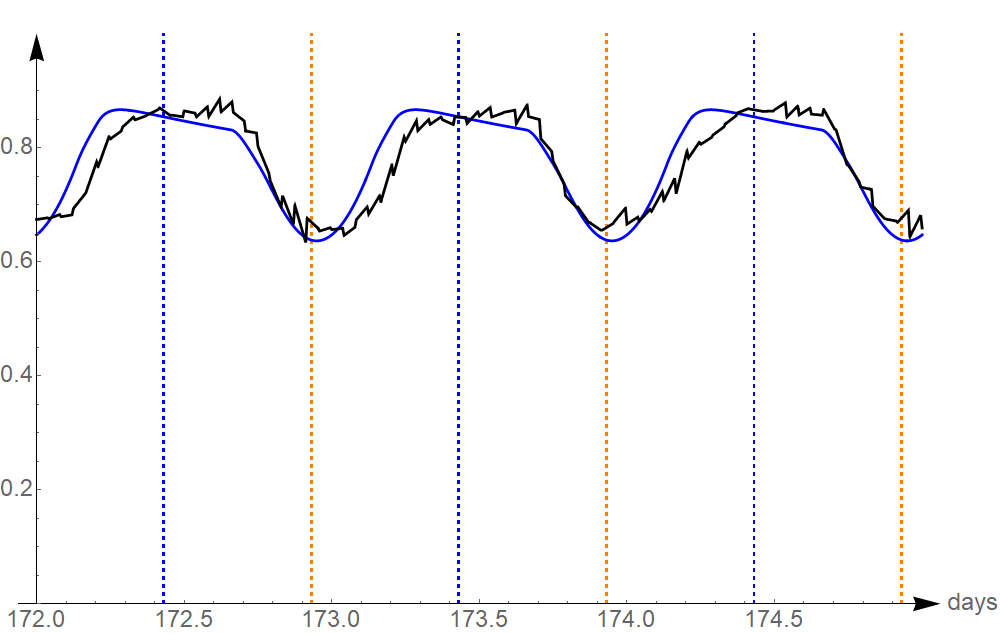}
\includegraphics[width = 0.25 \textwidth]{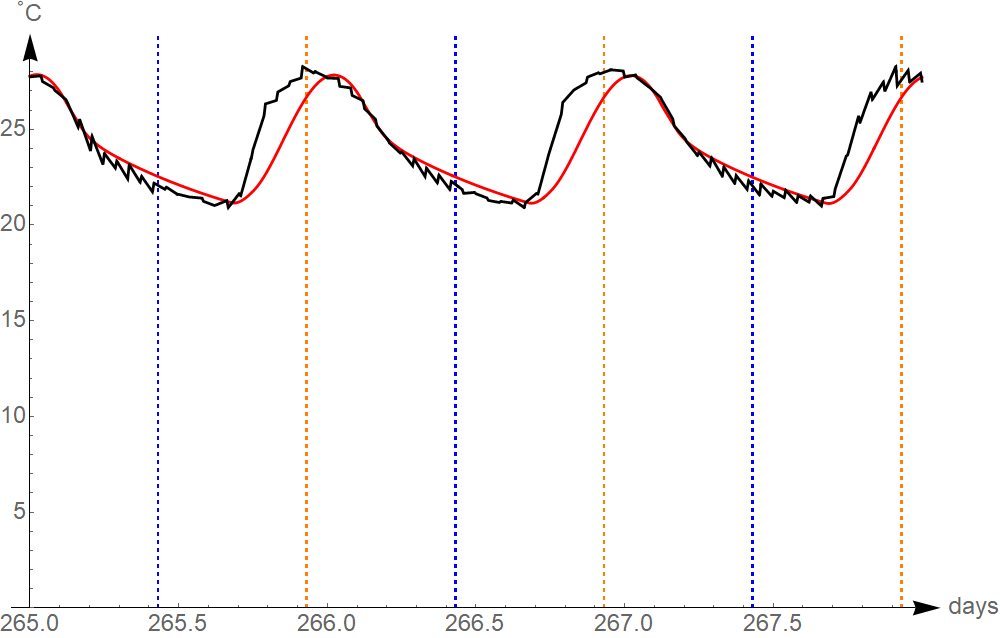}\includegraphics[width =0.25 \textwidth]{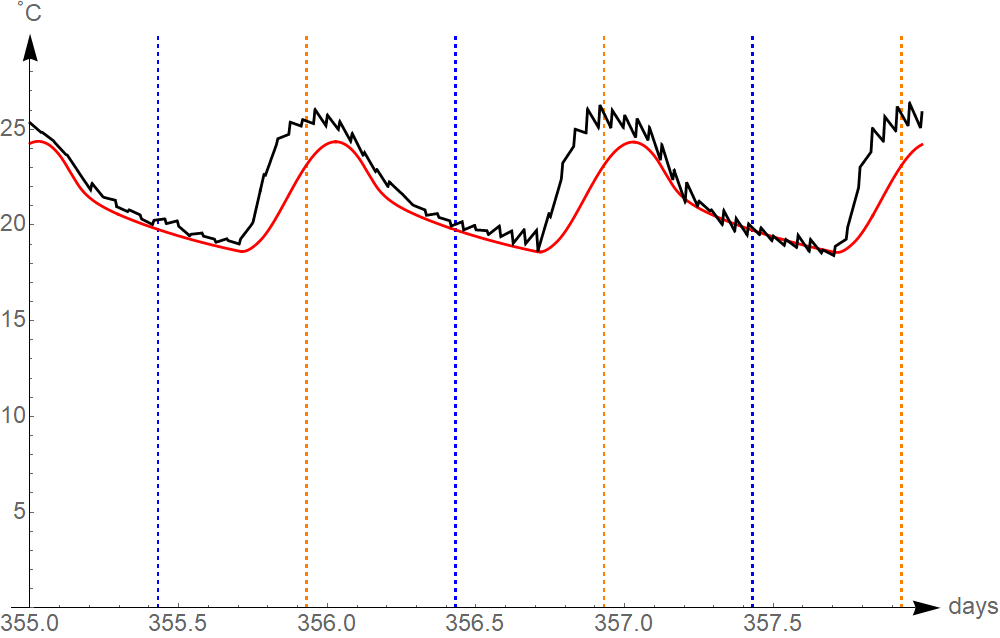}\includegraphics[width = 0.25 \textwidth]{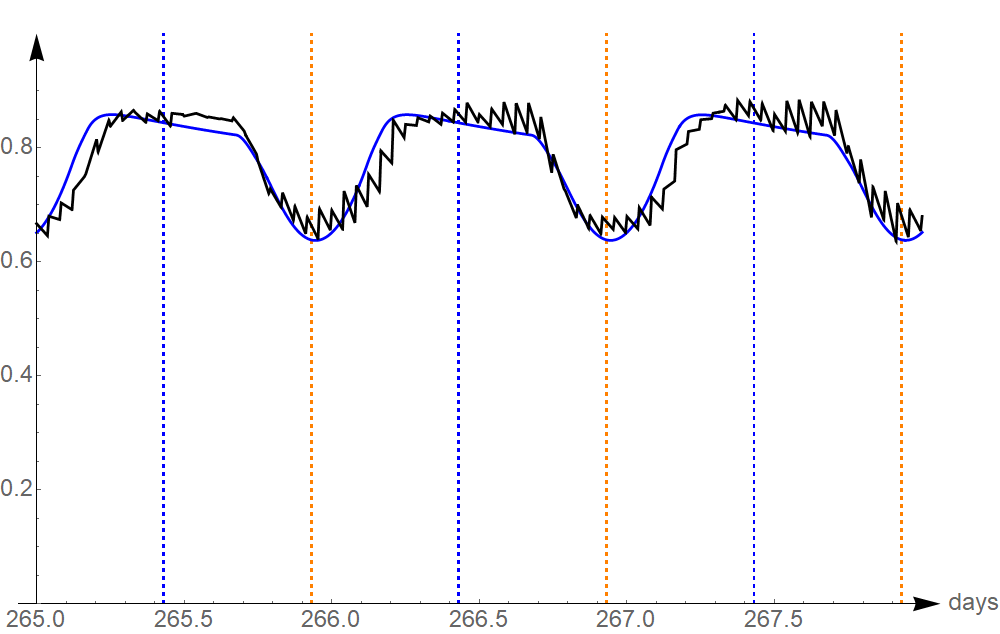}\includegraphics[width = 0.25 \textwidth]{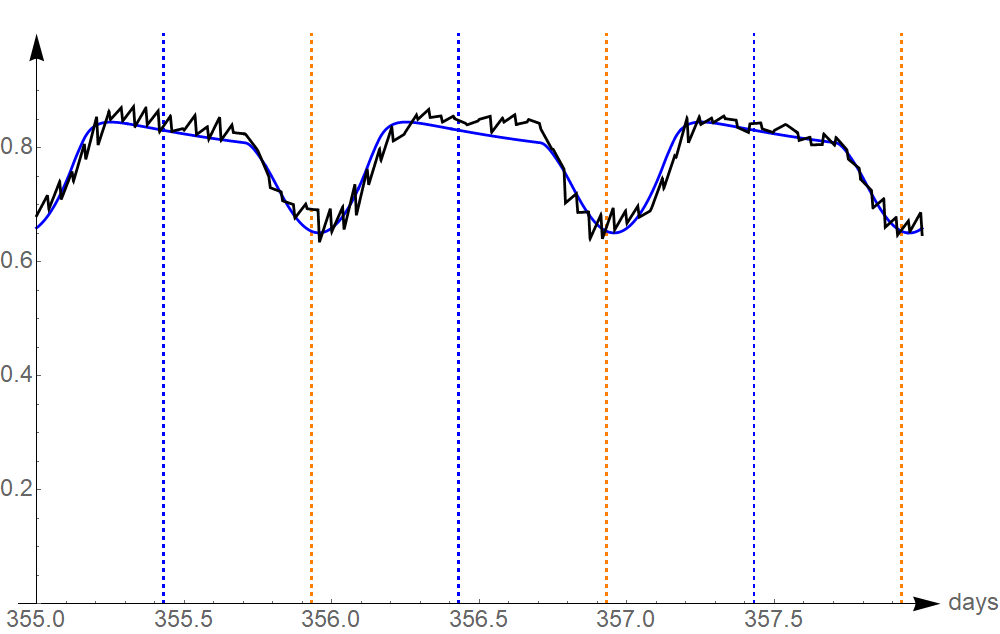}
\end{center}
\caption{Mean and computed temperature (left) and relative humidity (right) in Hilo during the year (top) and on solstices and equinoxes (bottom). In black the average temperatures and humidities, in red simulated temperatures, in blue simulated humidity, the grid lines represent solstices, equinoxes, noons and midnights.}
\label{fig:hilo}
\end{figure}

Hilo belongs to a region with Tropical, Rainforest K\"oppen climate (AF type). It is situated at latitude $19.72$ and longitude $-155.05$.
Belonging to an Hawaiian island, we choose $p = 0.1$. The presence of forest makes it reasonable to choose an higher value for the land's thermal capacity $C_1 = 1.7\times 10^6 \si{\joule \kelvin^{-1} \meter^{-2}}$, while considering $\ell_2 = 50\si{\meter}$ for oceans gives $C_2 = 2.1 \times 10^8 \si{\joule\kelvin^{-1}\meter^{-2}}$. We also consider the following values for other location-dependent parameters:
\[
\alpha_0^T = 0.8, \quad h_{02} = 25, \quad \mu_1 = 5.7\times 10^{-5}, \quad \nu = 1.7\times 10^{-5}.
\]
In Figure~\ref{fig:hilo} are represented the computed evolution of temperature (red) and humidity (blue) of the air and the real averaged temperatures and humidities (black) from 1973 to 2019.

\subsection{Arid climate: Kufra, Libya}
\begin{figure}[h]
\begin{center}
\includegraphics[width = 0.5 \textwidth]{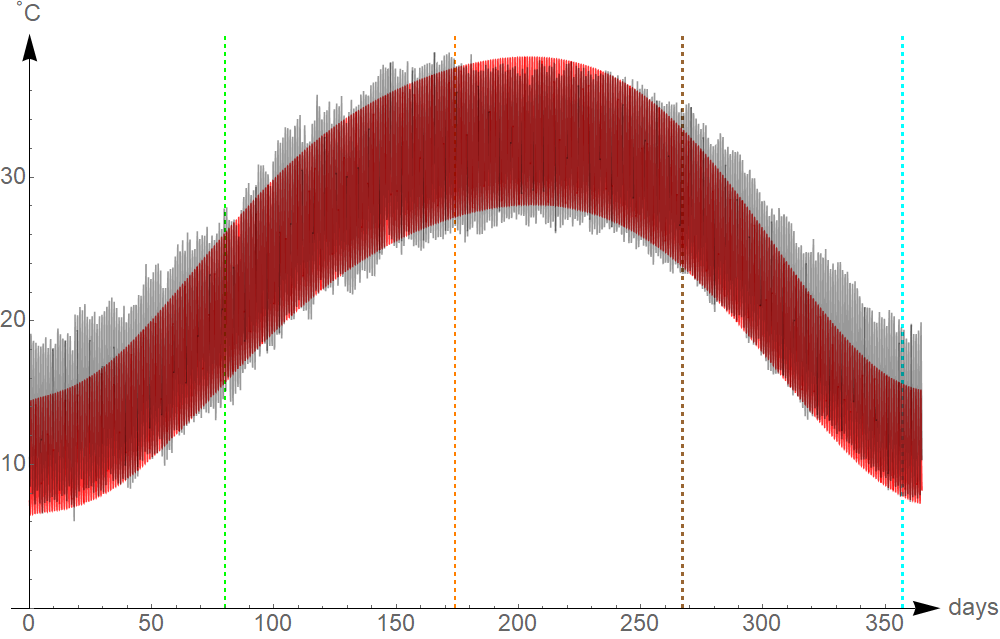}\includegraphics[width = 0.5 \textwidth]{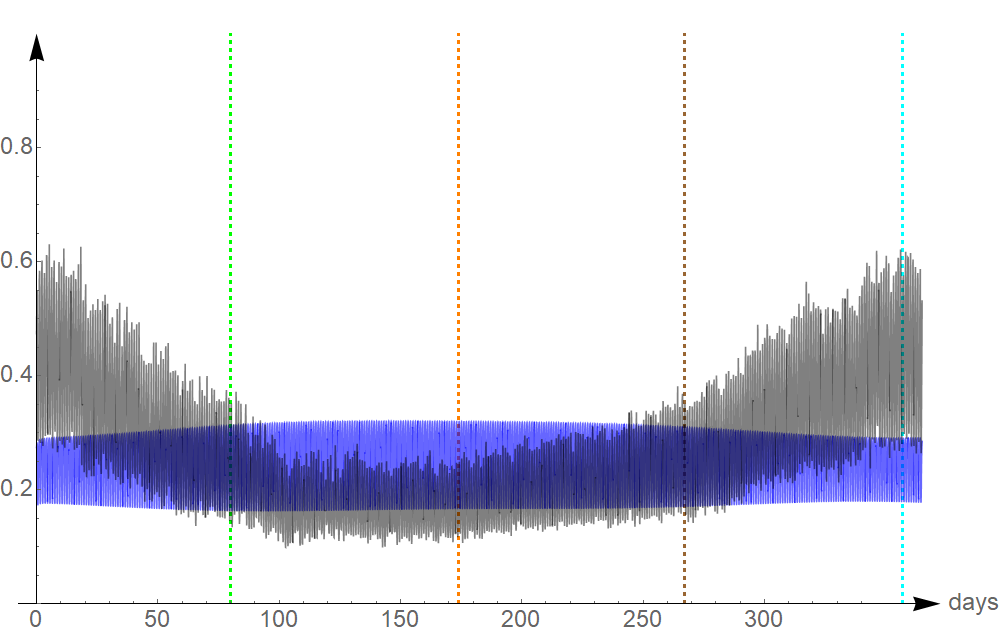}
\includegraphics[width = 0.25 \textwidth]{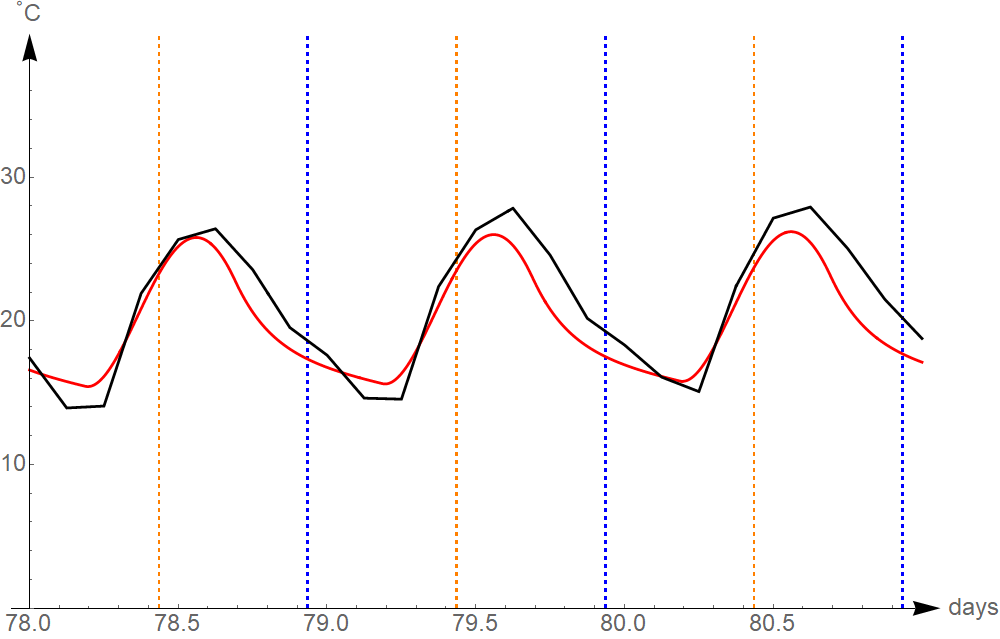}\includegraphics[width = 0.25 \textwidth]{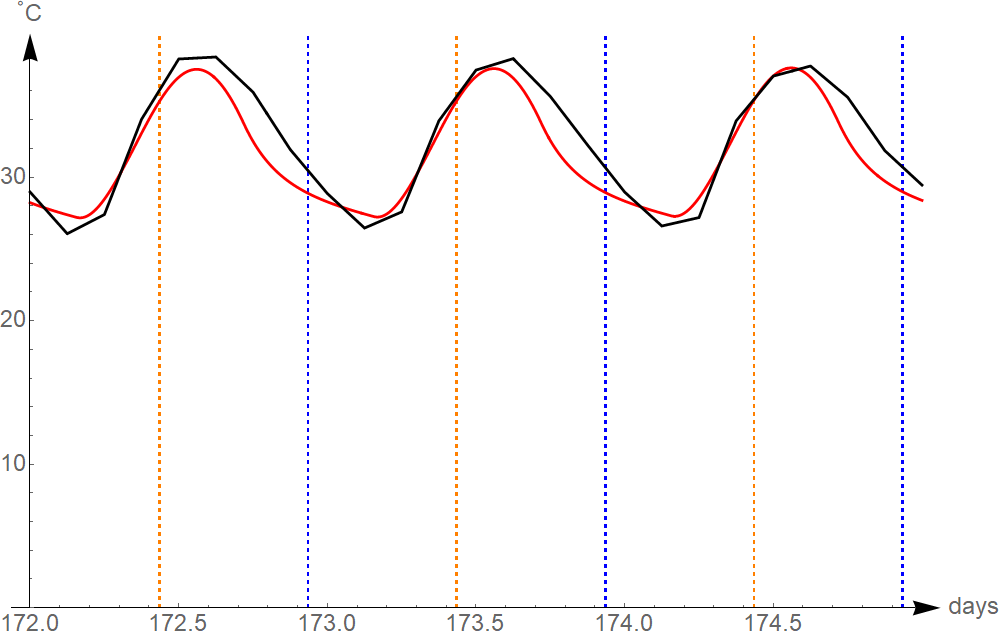}\includegraphics[width = 0.25 \textwidth]{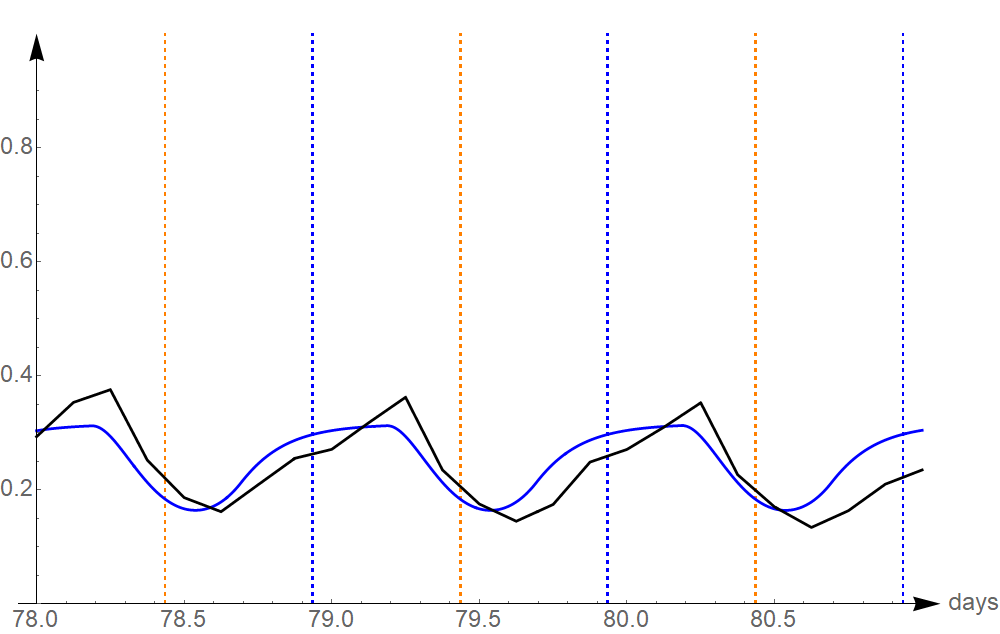}\includegraphics[width = 0.25 \textwidth]{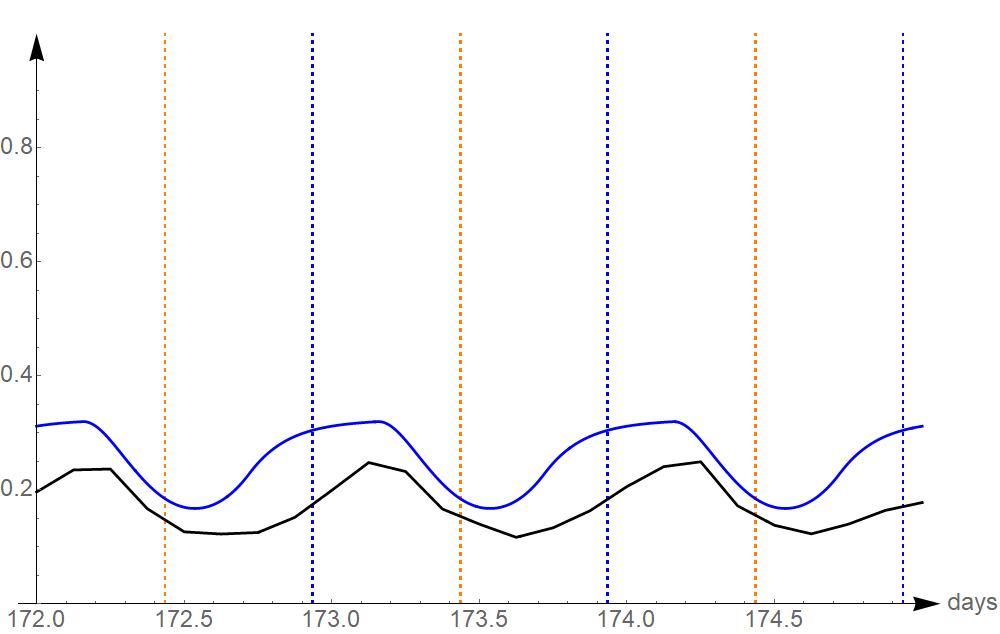}
\includegraphics[width = 0.25 \textwidth]{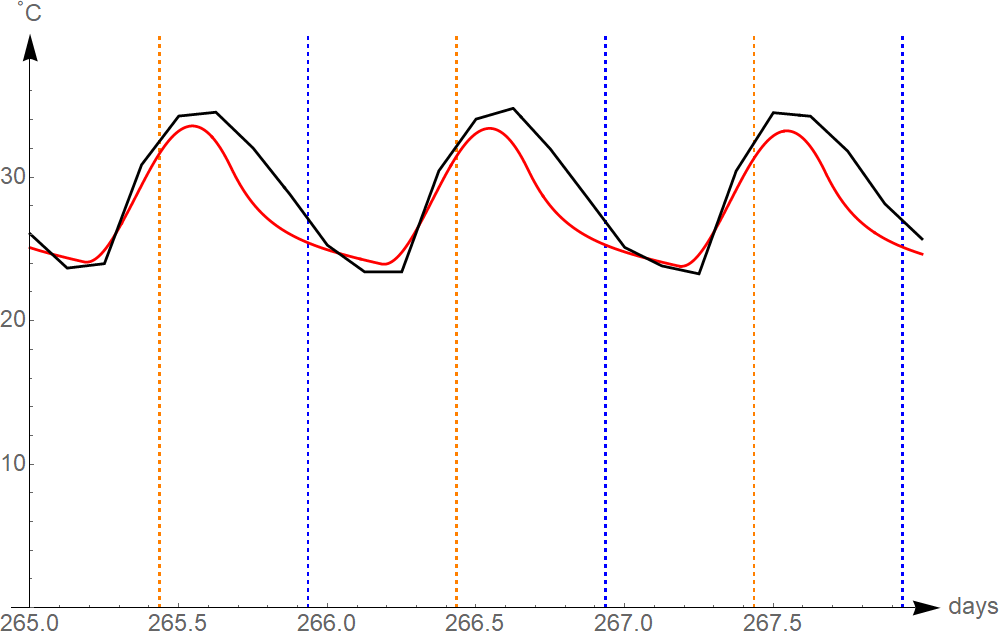}\includegraphics[width =0.25 \textwidth]{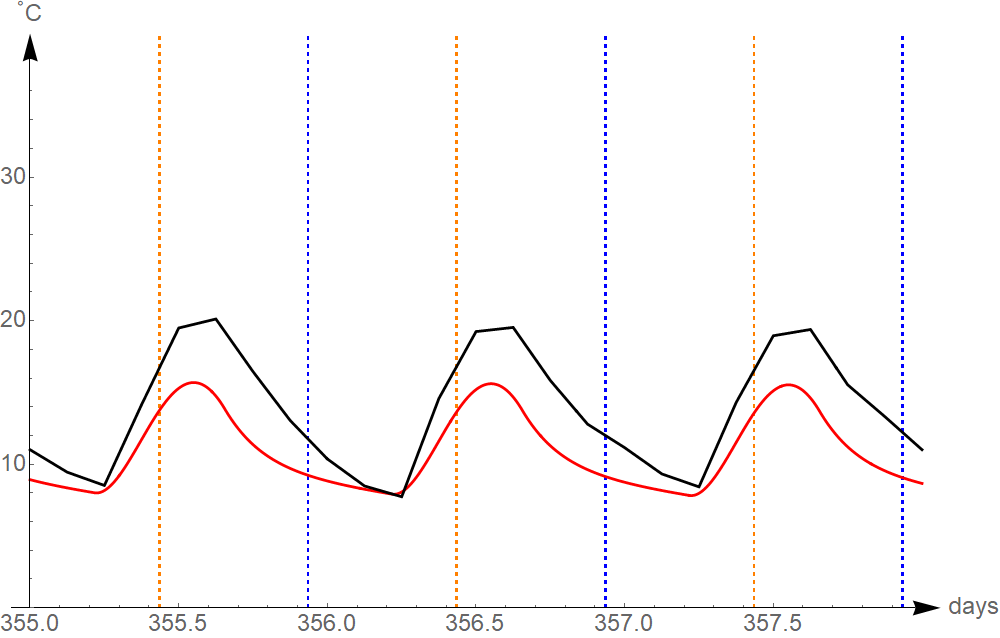}\includegraphics[width = 0.25 \textwidth]{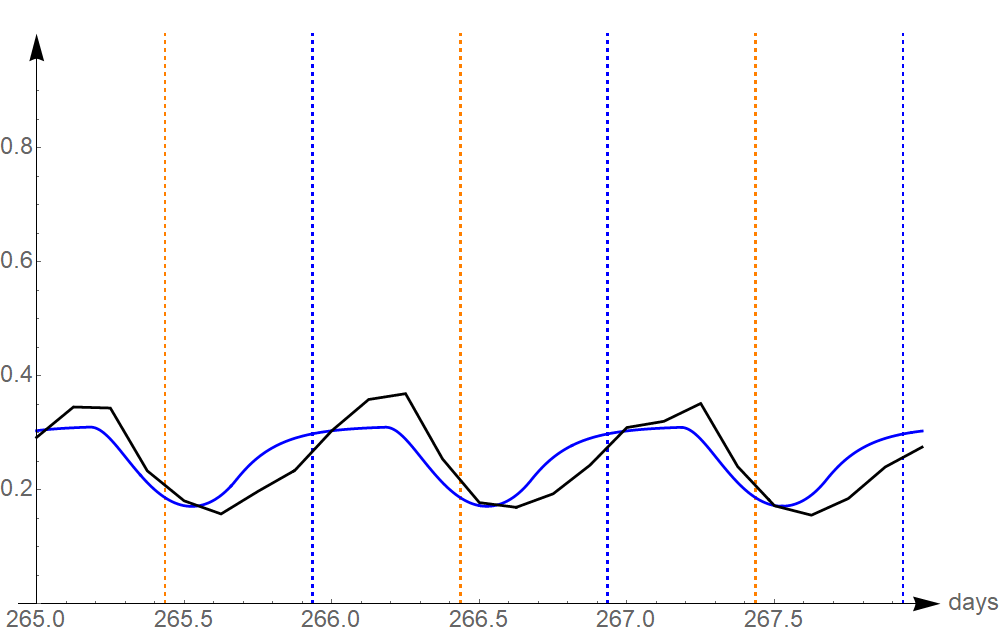}\includegraphics[width = 0.25 \textwidth]{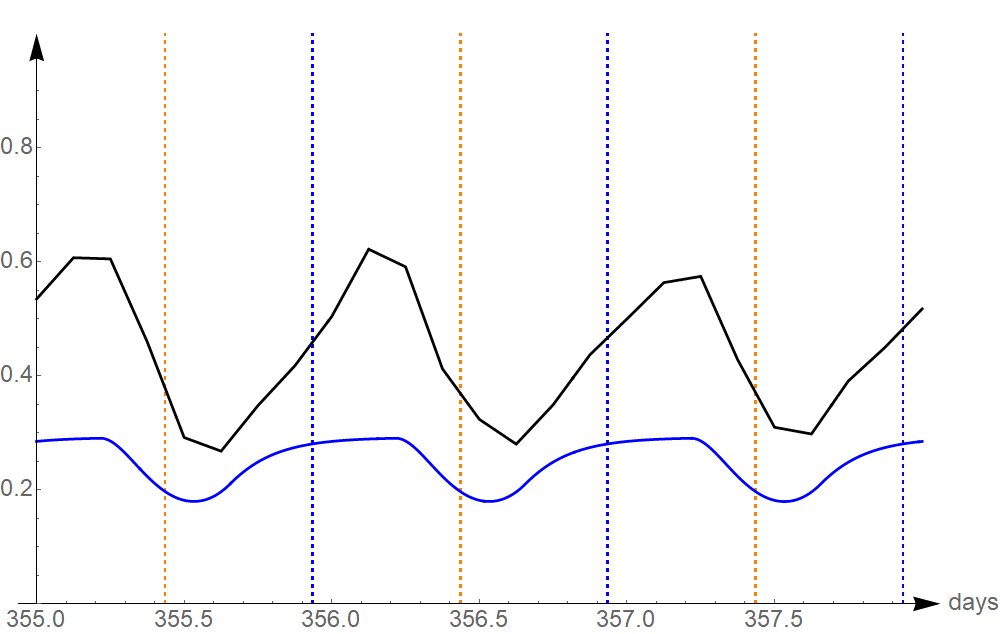}
\caption{Mean and computed temperature (left) and relative humidity (right) in Kufra during the year (top) and on solstices and equinoxes (bottom). In black the average temperatures and humidities, in red simulated temperatures, in blue simulated humidity, the grid lines represent solstices, equinoxes, noons and midnights.}
\label{fig:kufra}
\end{center}
\end{figure}

Kufra belongs to the eastern part of Sahara with Arid, Hot Desert (BWH type) K\"oppen climate. It is situated at latitude 24.18 and longitude 23.31. Being in a desert, water has almost no influence and we hence have chosen $p = 0.9$. We recall the choices
\[
C_1 = 3.2 \times 10^6, \quad \ell_2 = 40, \quad \alpha_0^T = 0.88, \quad h_{02} = 30, \quad \mu_1 = 2.9\times 10^{-6}, \quad \nu = 10^{-5}.
\]
All other parameters are in the Tables, and we have used the coefficients for sand. 
In Figure~\ref{fig:kufra} are represented the computed evolution of temperature (red) and humidity (blue) of the air and the real averaged temperatures and humidities (black) from 1973 to 2019.

\subsection{Temperate climate: Catania, Italy}
Catania is one of the cities on the Mediterranean Sea with Temperate, Hot-summer, Mediterranean K\"oppen climate (CSA type). It is situated at latitude 37.47 and longitude $15.05$. Given its location, we choose $p = 0.6$. Considering that the top layer of Mediterranean sea mix to a depth of up to $40\si{\meter} = \ell_2$, we consider
\[
\alpha_0^T = 0.85, \quad h_{02} = 35, \quad \mu_1 = 5.7 \times 10^{-5}, \quad \nu = 2 \times 10^{-5}.
\]

\begin{figure}[h!]
\begin{center}
\includegraphics[width = 0.5 \textwidth]{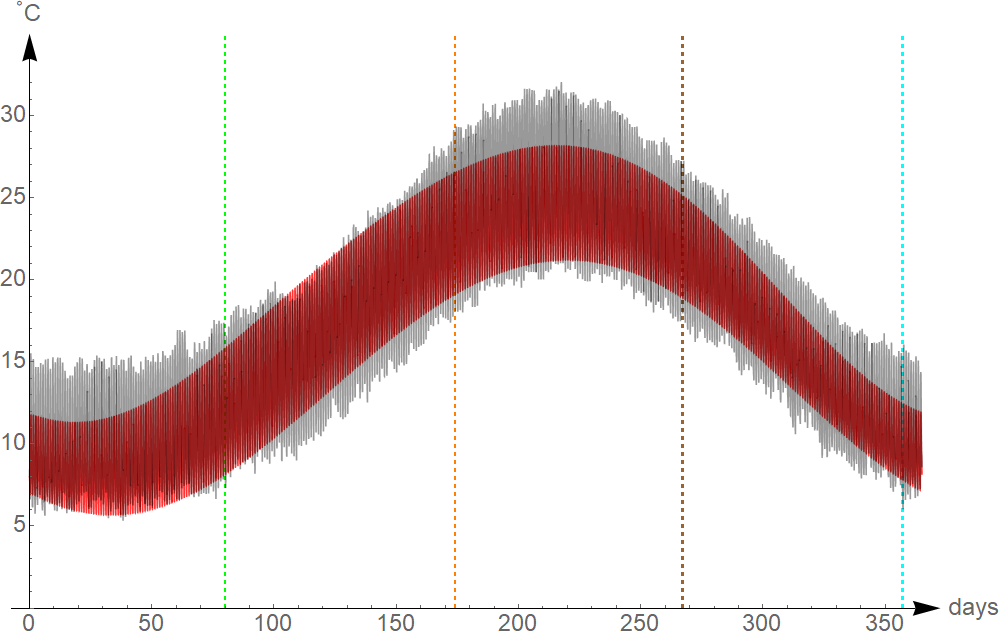}\includegraphics[width = 0.5 \textwidth]{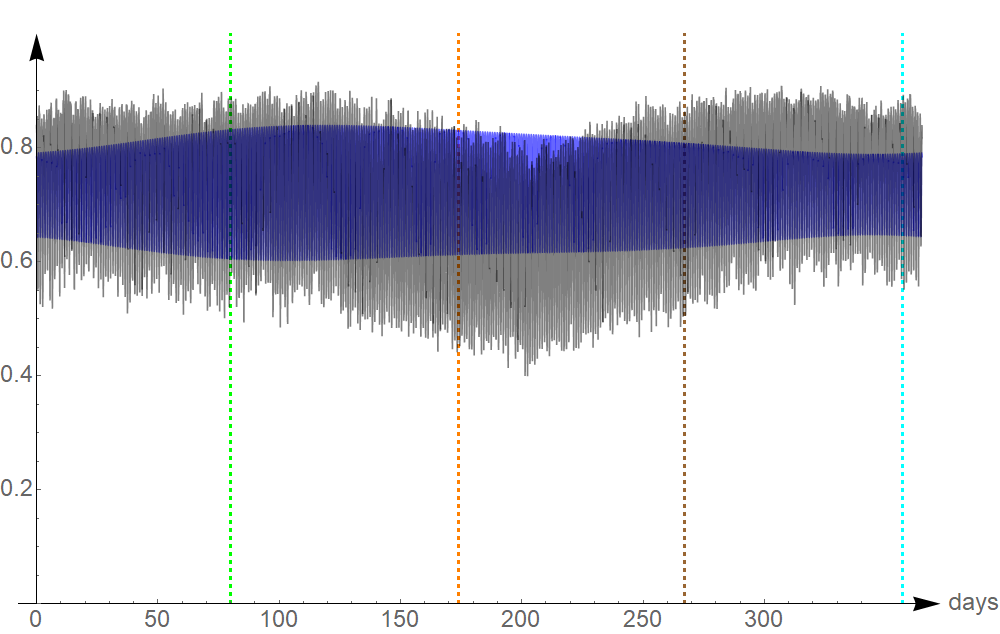}
\includegraphics[width = 0.25 \textwidth]{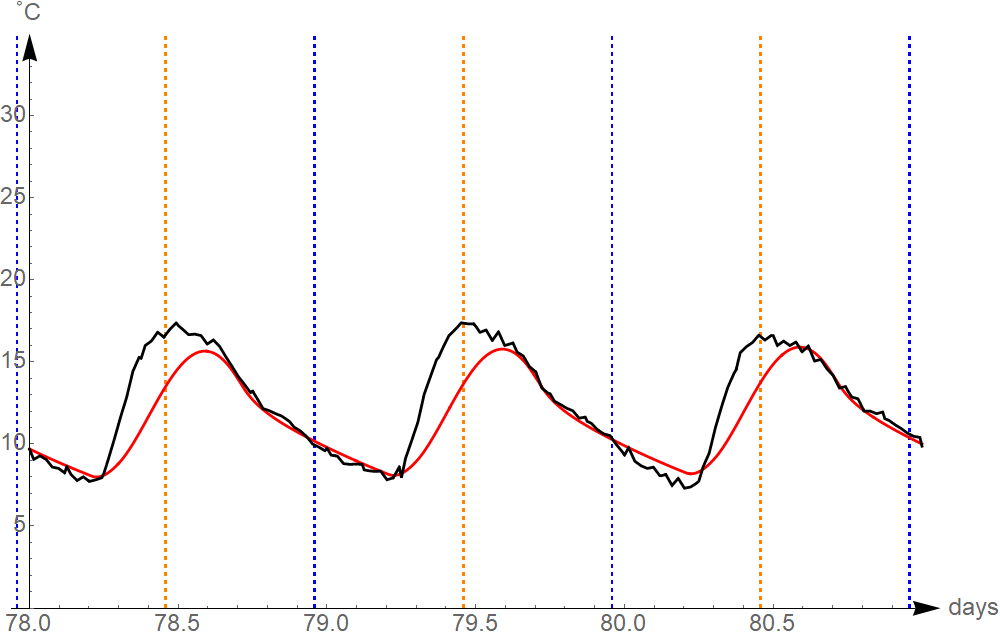}\includegraphics[width = 0.25 \textwidth]{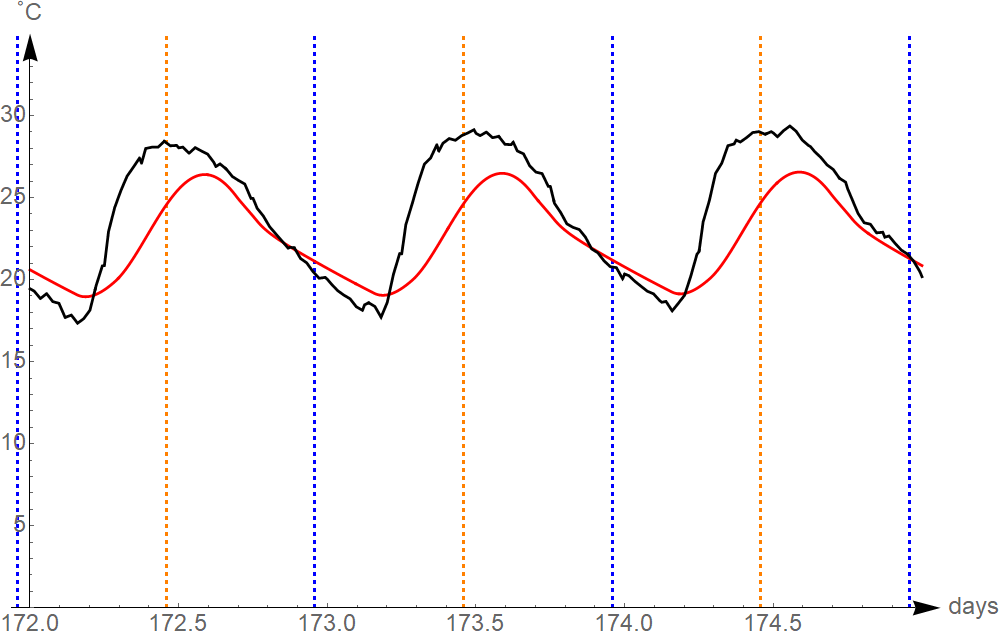}\includegraphics[width = 0.25 \textwidth]{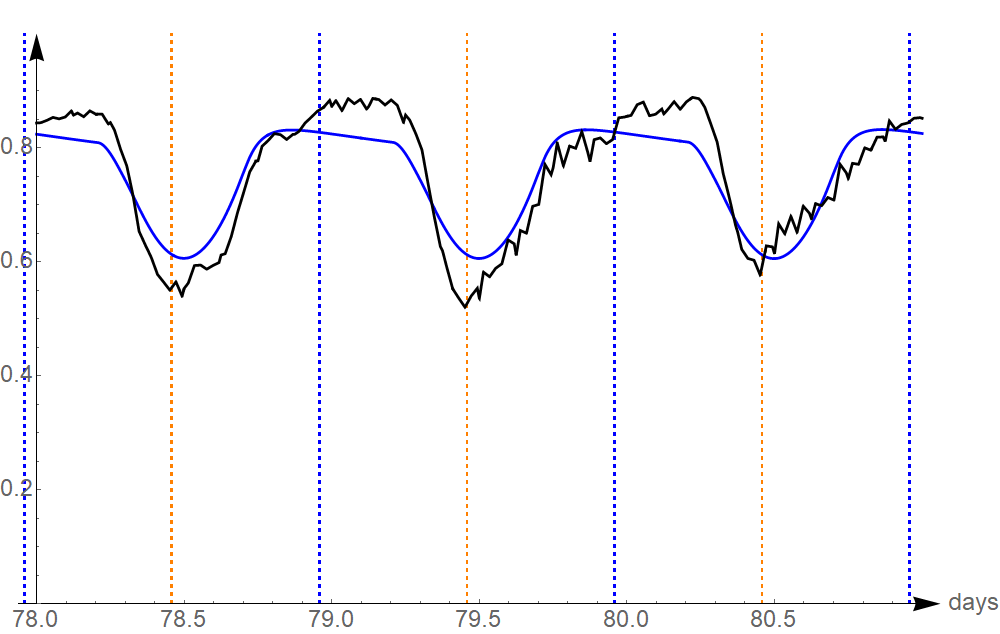}\includegraphics[width = 0.25 \textwidth]{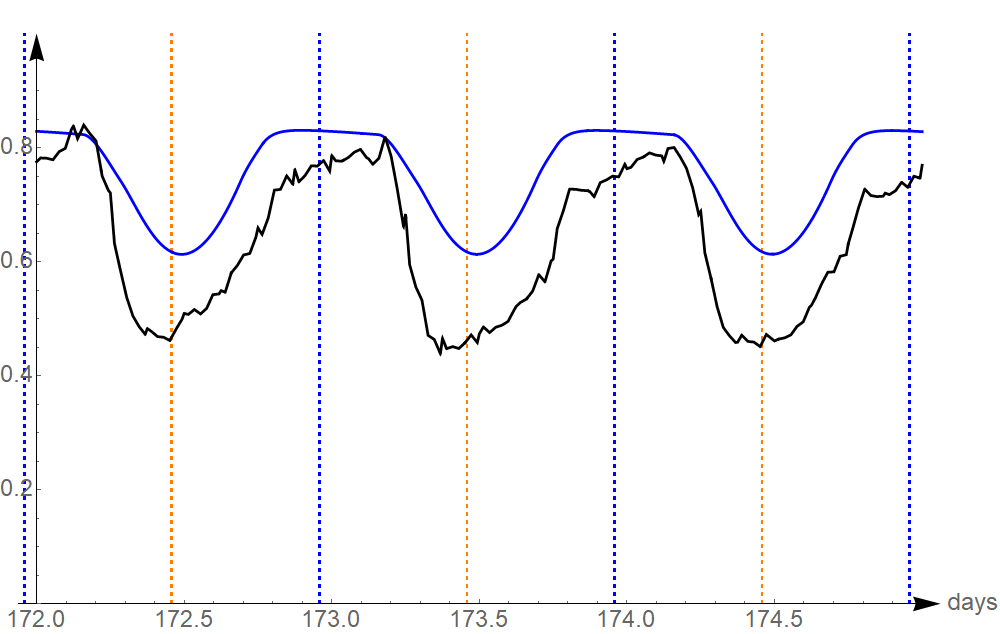}
\includegraphics[width = 0.25 \textwidth]{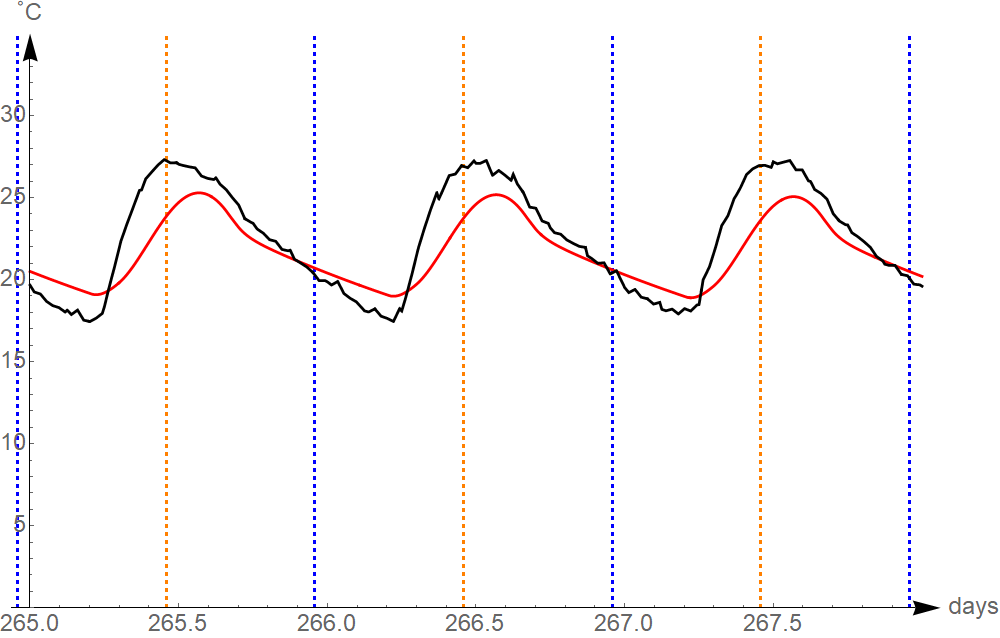}\includegraphics[width =0.25 \textwidth]{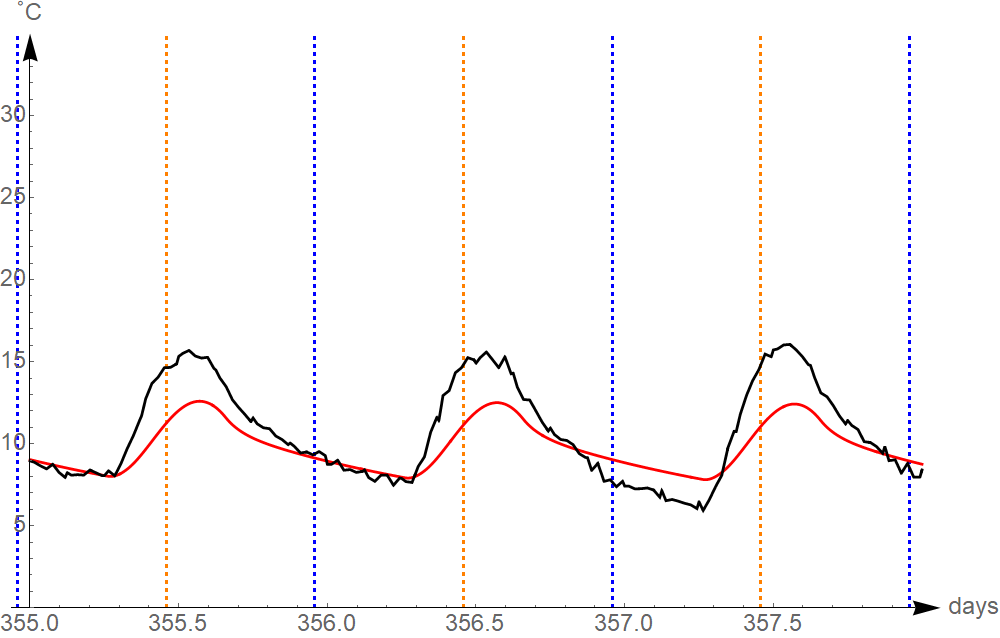}\includegraphics[width = 0.25 \textwidth]{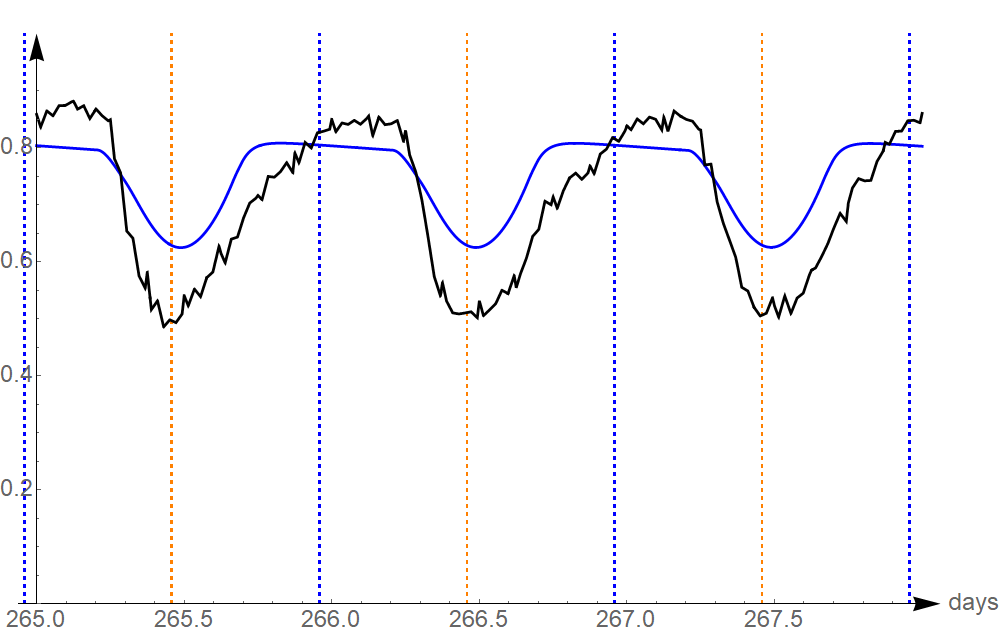}\includegraphics[width = 0.25 \textwidth]{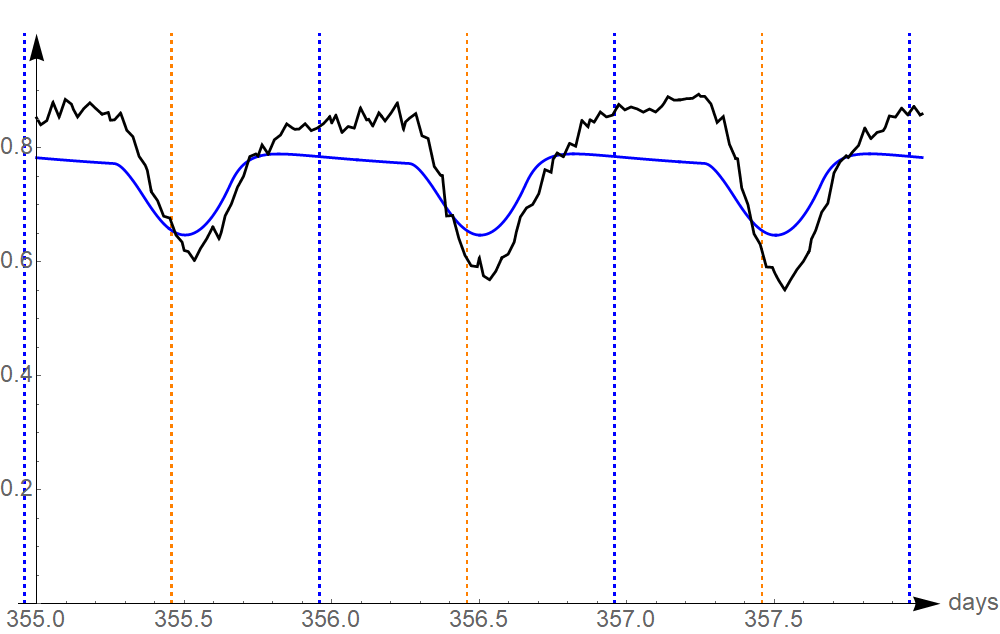}
\caption{Mean and computed temperature (left) and relative humidity (right) in Catania during the year (top) and on solstices and equinoxes (bottom). In black the average temperatures and humidities, in red simulated temperatures, in blue simulated humidity, the grid lines represent solstices, equinoxes, noons and midnights.}
\label{fig:catania}
\end{center}
\end{figure}

In Figure~\ref{fig:catania} are represented the computed evolution of temperature (red) and humidity (blue) of the air and the real averaged temperatures and humidities (black) from 1973 to 2019.

\subsection{Continental climate: Lincoln, USA}
Lincoln belongs to the central USA, a region with Continental, Hot-summer, Humid K\"oppen climate (DFA type). It is situated at latitude $40.85$ and longitude $-96.75$. For its location, we choose $p = 0.8$, $\ell_2 = 40\si{\meter}$. We adopt
\[
\alpha_0^T = 0.84, \quad h_{02} = 35, \quad \mu_1 = 4.6\times 10^{-5}, \nu = 2 \times 10^{-5}
\]

\begin{figure}[h!]
\begin{center}
\includegraphics[width = 0.5 \textwidth]{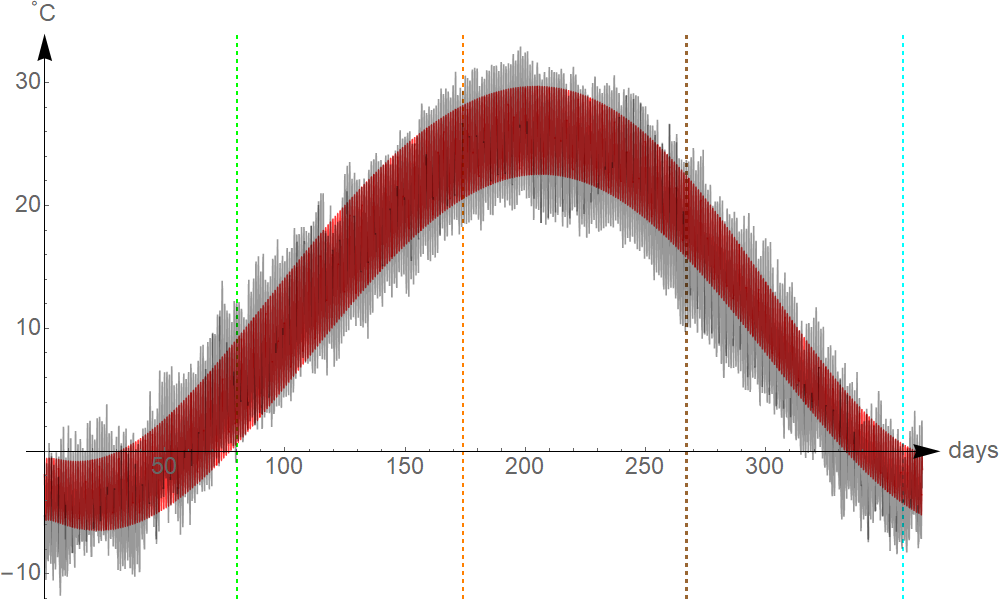}\includegraphics[width = 0.5 \textwidth]{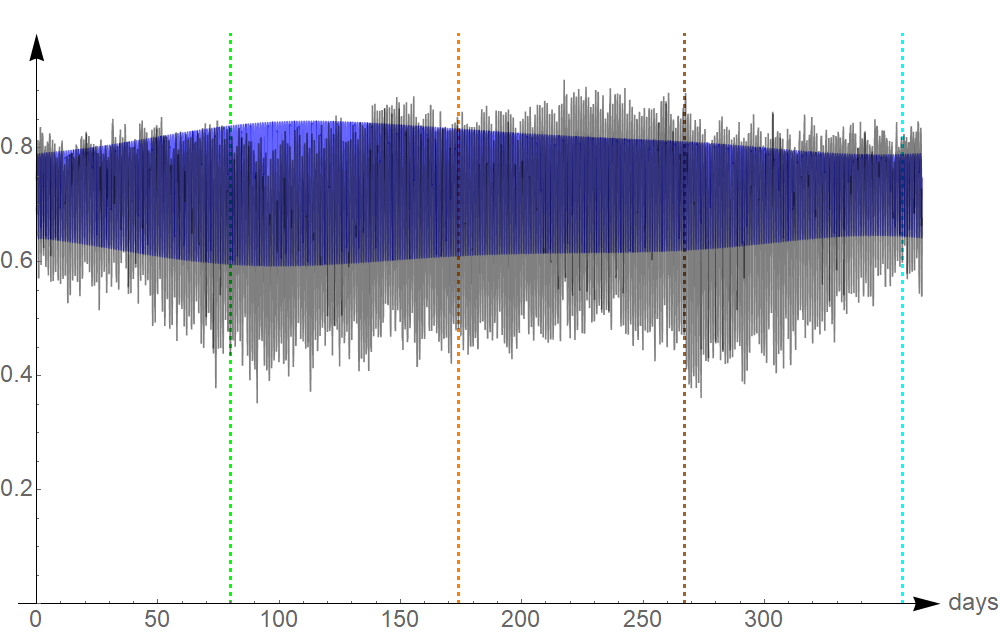}
\includegraphics[width = 0.25 \textwidth]{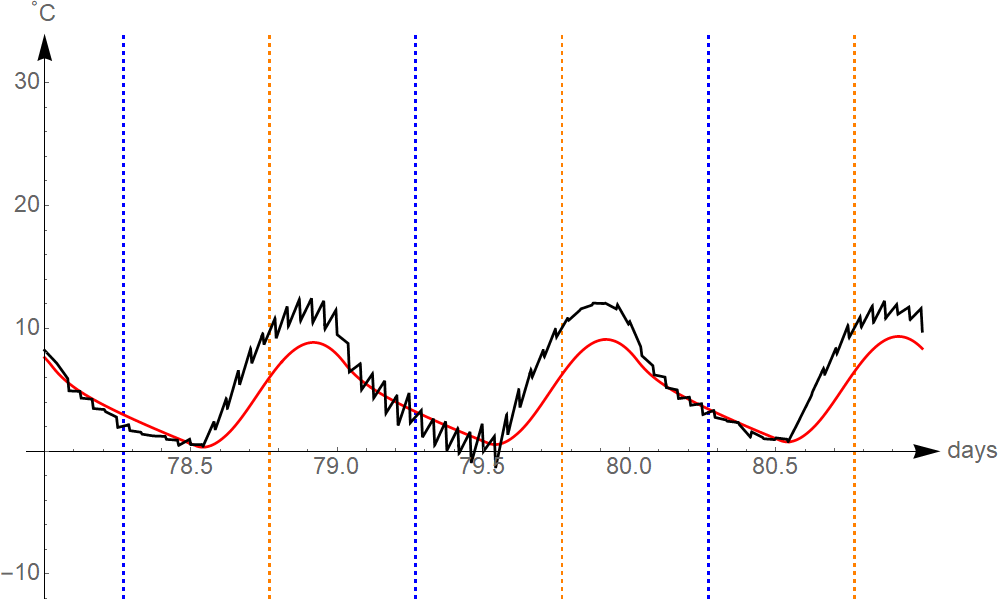}\includegraphics[width = 0.25 \textwidth]{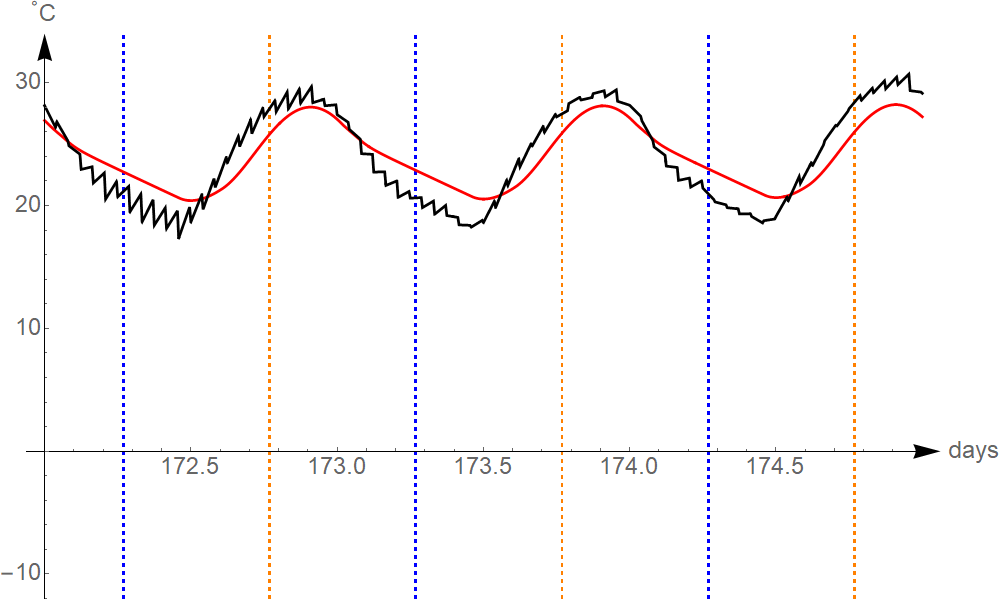}\includegraphics[width = 0.25 \textwidth]{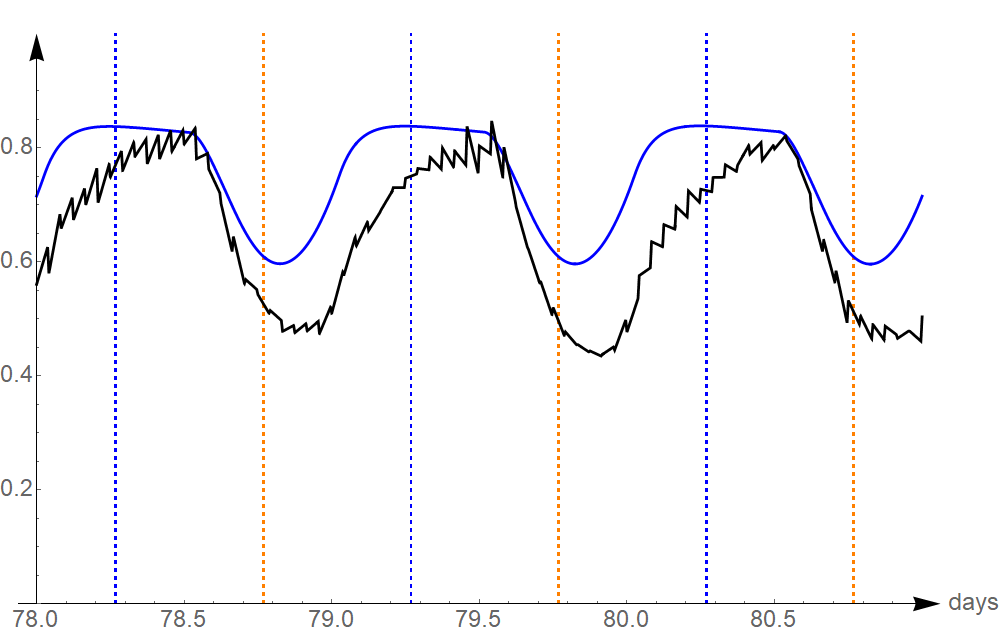}\includegraphics[width = 0.25 \textwidth]{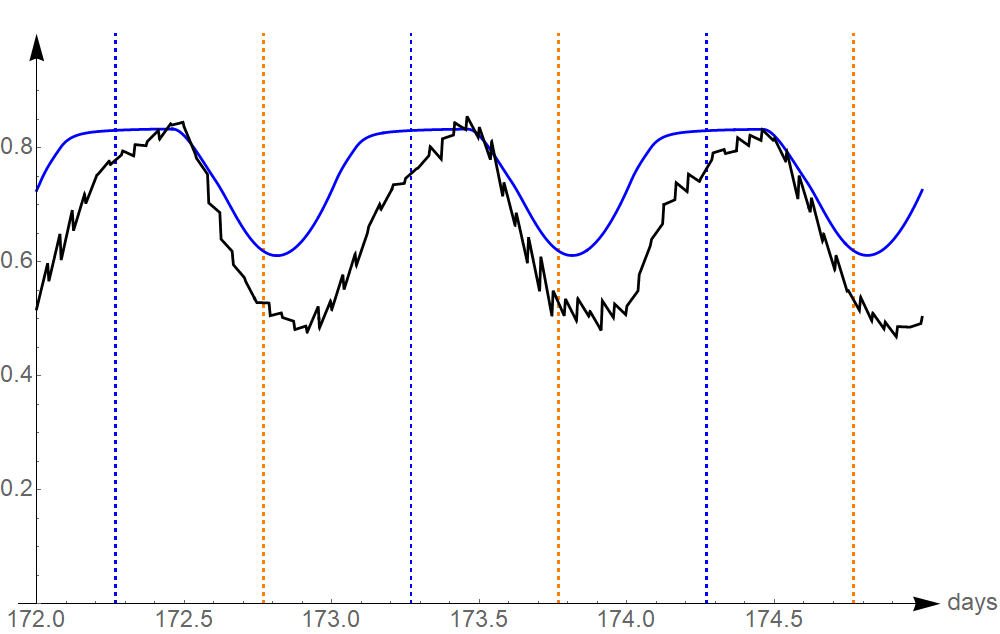}
\includegraphics[width = 0.25 \textwidth]{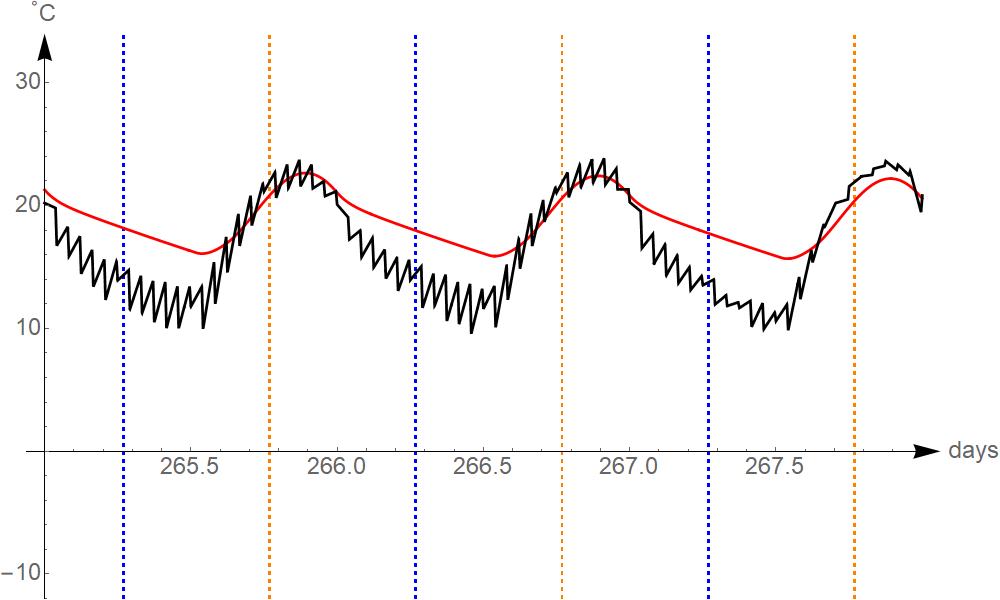}\includegraphics[width =0.25 \textwidth]{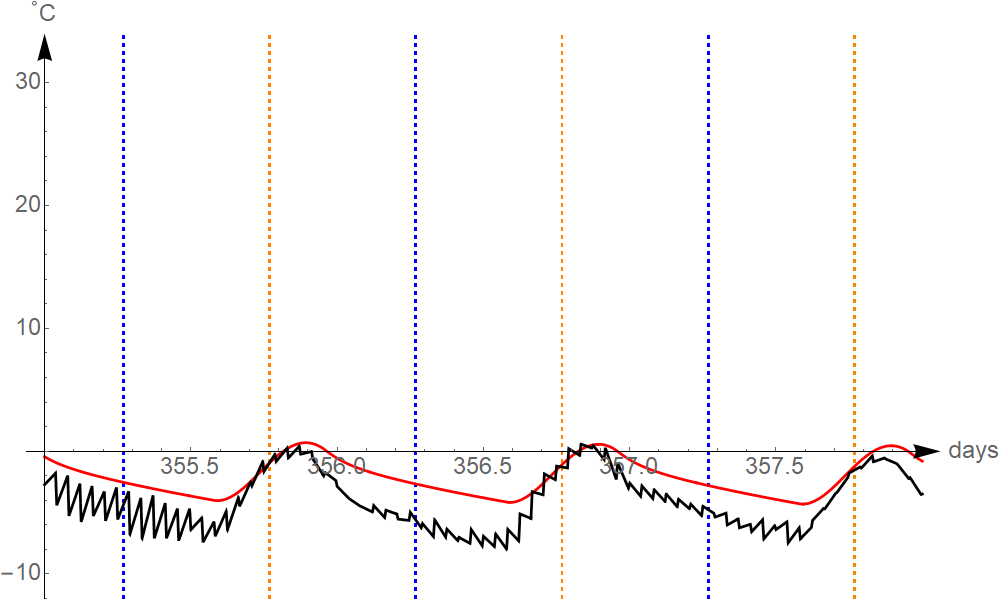}\includegraphics[width = 0.25 \textwidth]{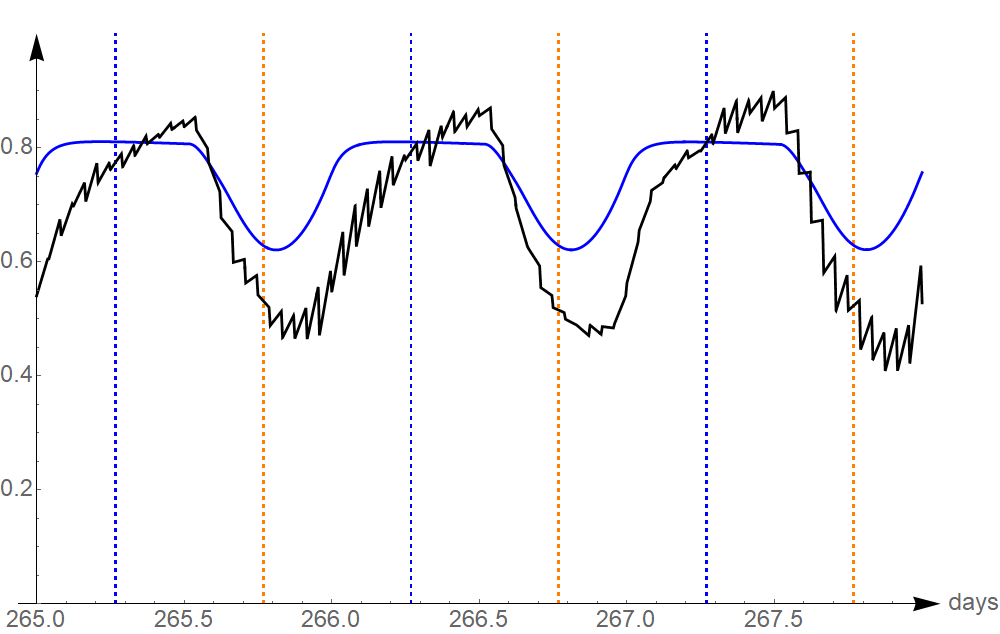}\includegraphics[width = 0.25 \textwidth]{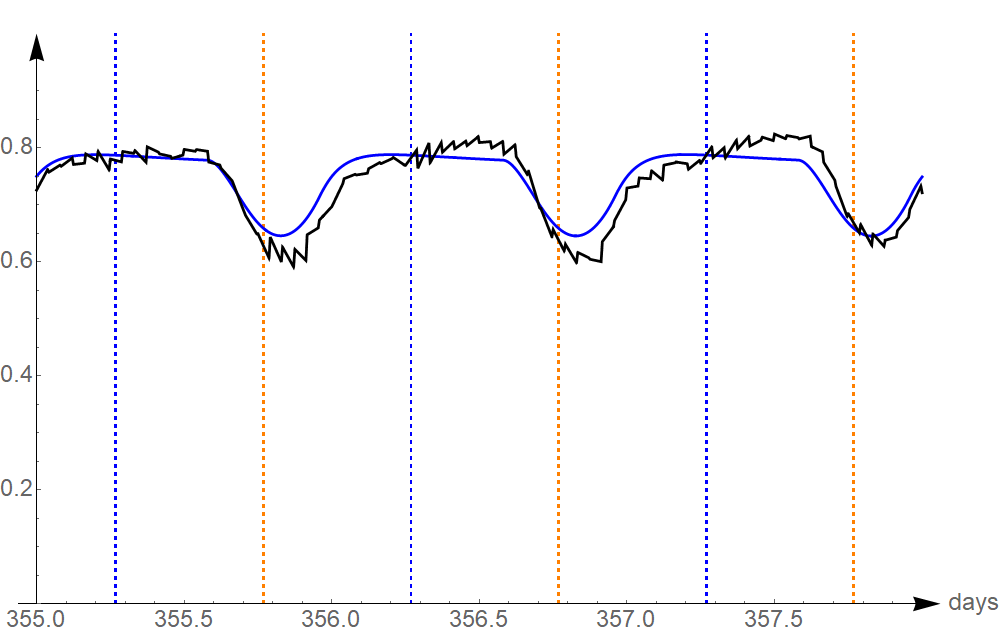}
\caption{Mean and computed temperature (left) and relative humidity (right) in Lincoln during the year (top) and on solstices and equinoxes (bottom). In black the average temperatures and humidities, in red simulated temperatures, in blue simulated humidity, the grid lines represent solstices, equinoxes, noons and midnights.}
\label{fig:lincoln}
\end{center}
\end{figure}

In Figure~\ref{fig:lincoln} are represented the computed evolution of temperature (red) and humidity (blue) of the air and the real averaged temperatures and humidities (black) from 1973 to 2019.

\subsection{Polar climate: Vostok, Antarctica}
Vostok is a weather station close to a lake in Antartica, it is located almost at the South Pole and it has Polar, Ice cap K\"oppen climate (EF type). This region is situated at latitude $-78.45$ and longitude $106.87$ and is always covered with ice and snow, living in eternal winter. The thermal inertia of the ice cap is very high and so, even if located on land, we have chosen $p = 0.45$,
\[
\alpha_0^T = 0.75, \quad h_{02} = 6, \quad \mu_1 =10^{-4}, \quad \nu = 5\times 10^{-5},
\]
All other parameters are in the Tables, and we have used the coefficients for ice over land and over water. 

\begin{figure}[h]
\begin{center}
\includegraphics[width = 0.5 \textwidth]{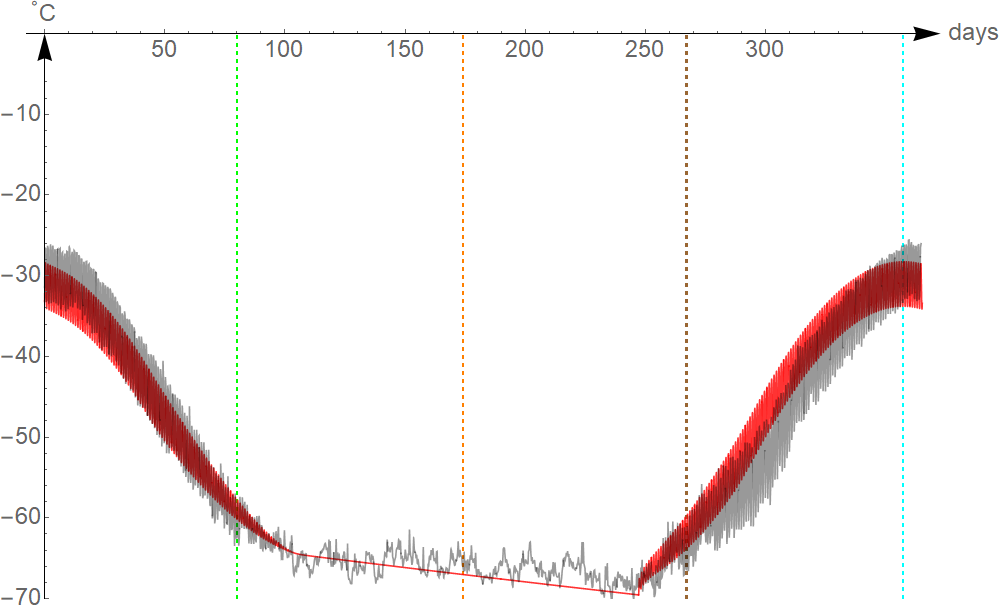}\includegraphics[width = 0.5 \textwidth]{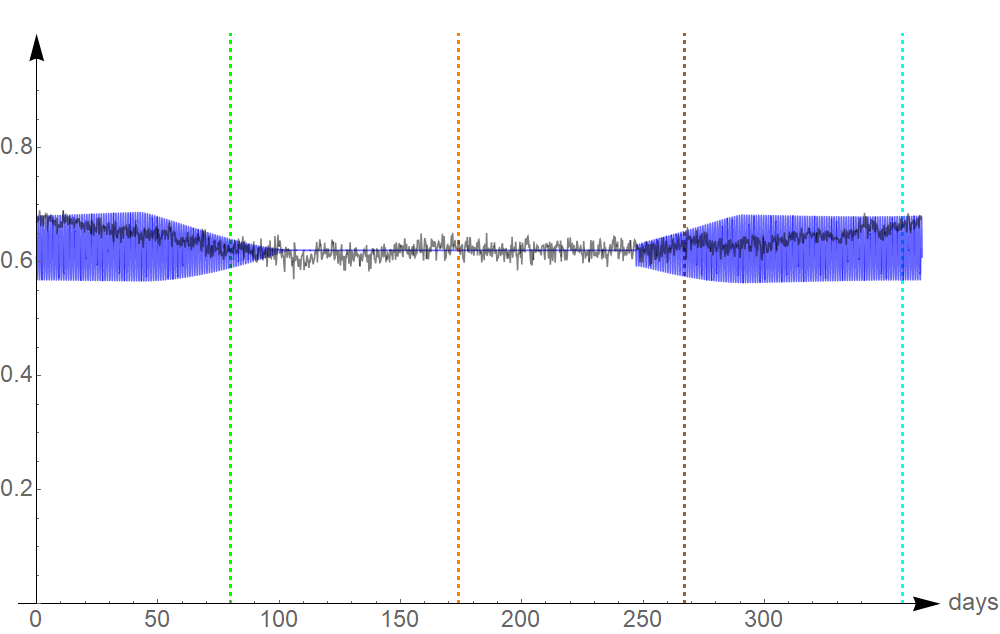}
\includegraphics[width = 0.25 \textwidth]{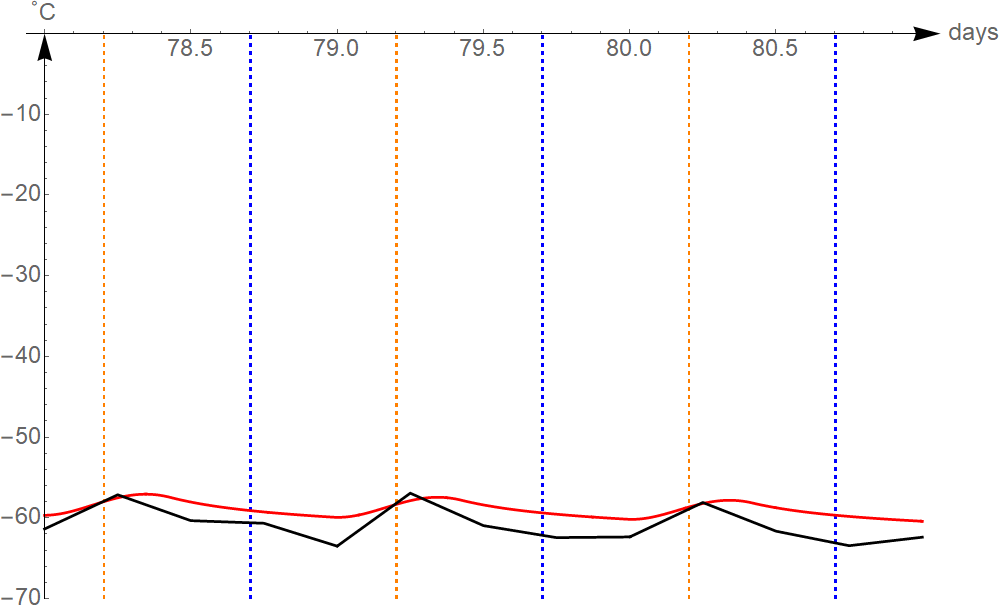}\includegraphics[width = 0.25 \textwidth]{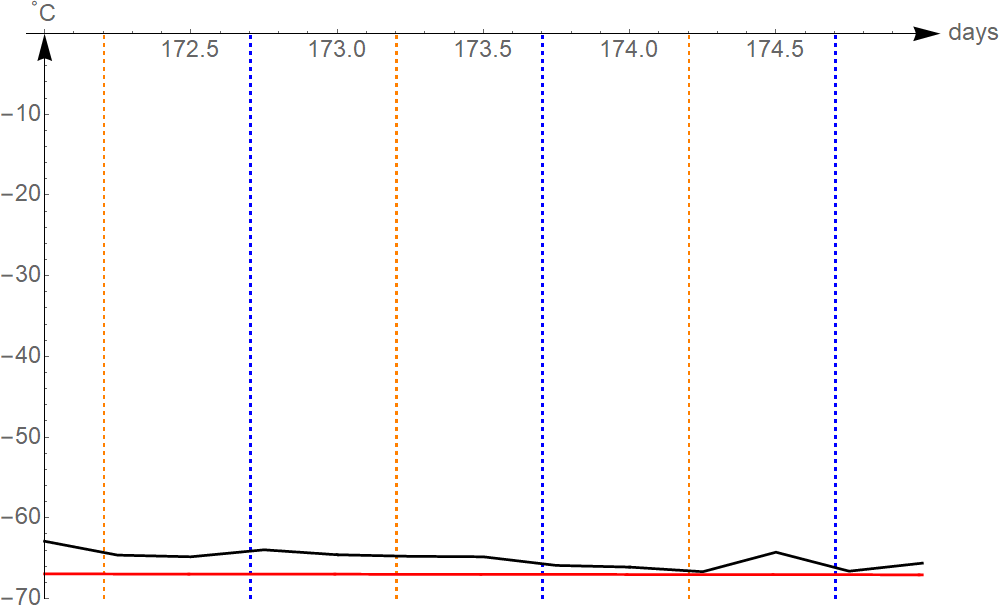}\includegraphics[width = 0.25 \textwidth]{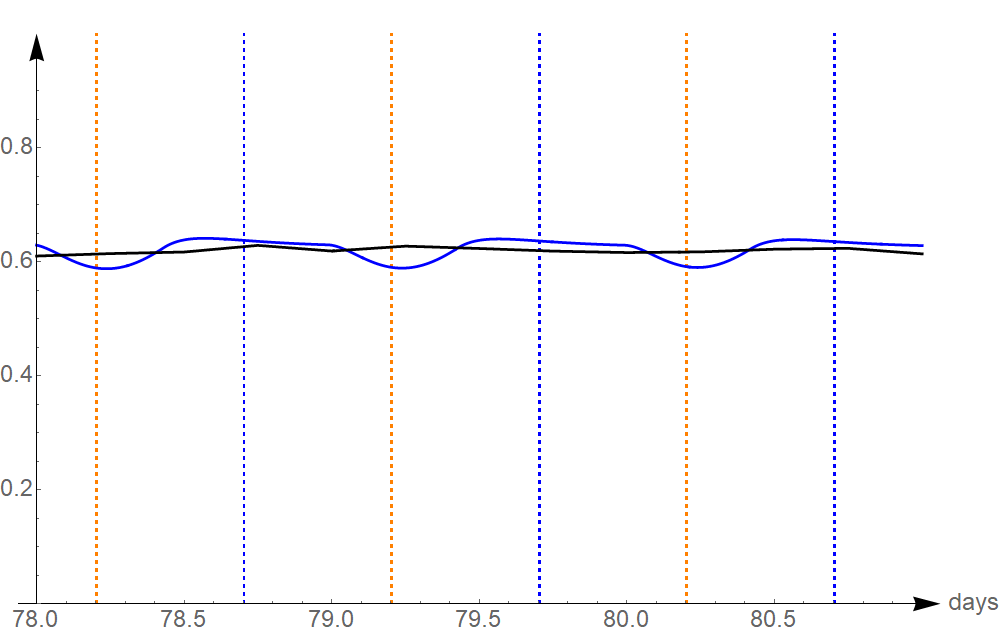}\includegraphics[width = 0.25 \textwidth]{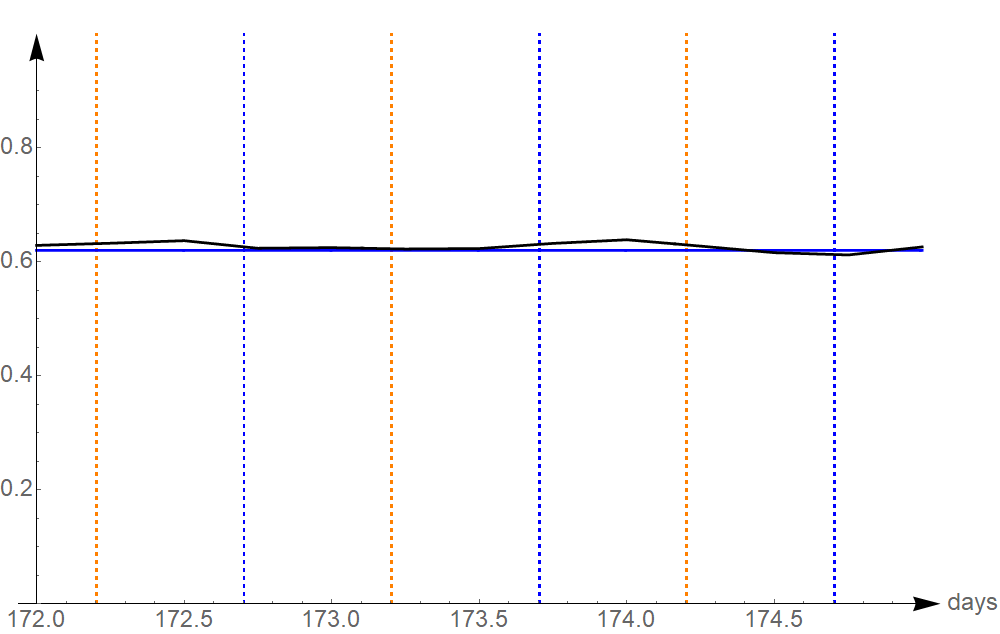}
\includegraphics[width = 0.25 \textwidth]{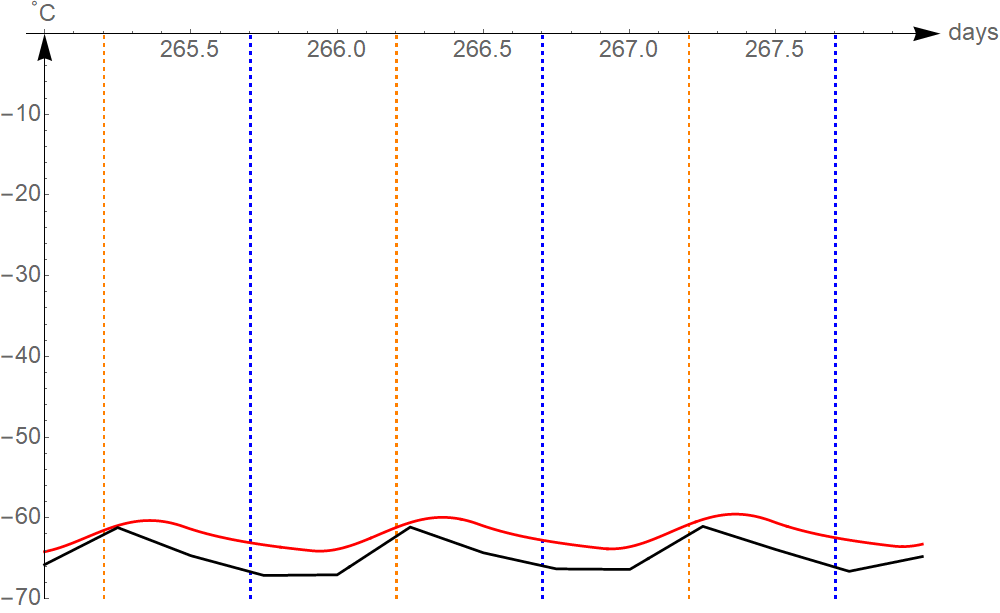}\includegraphics[width =0.25 \textwidth]{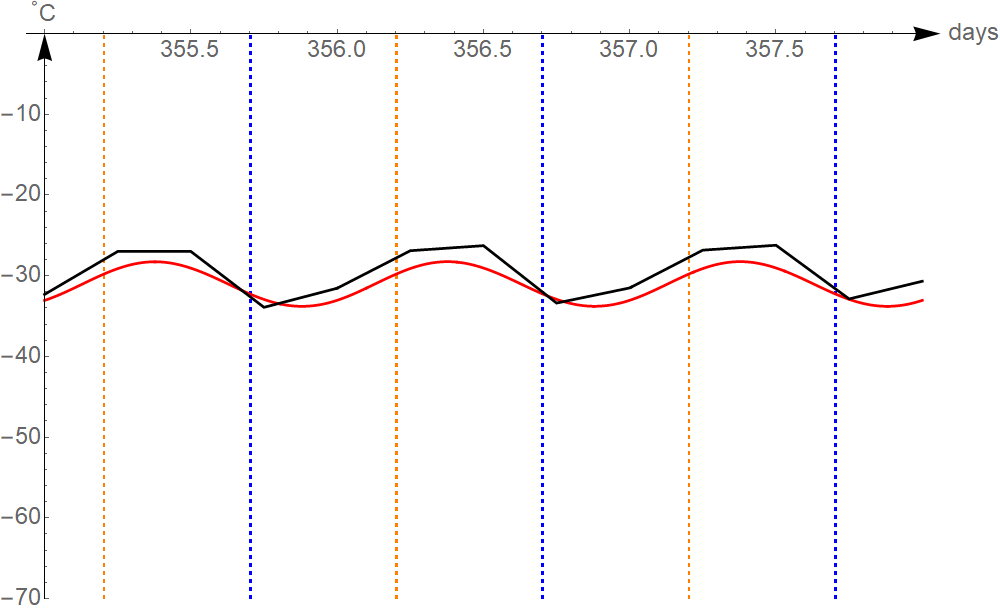}\includegraphics[width = 0.25 \textwidth]{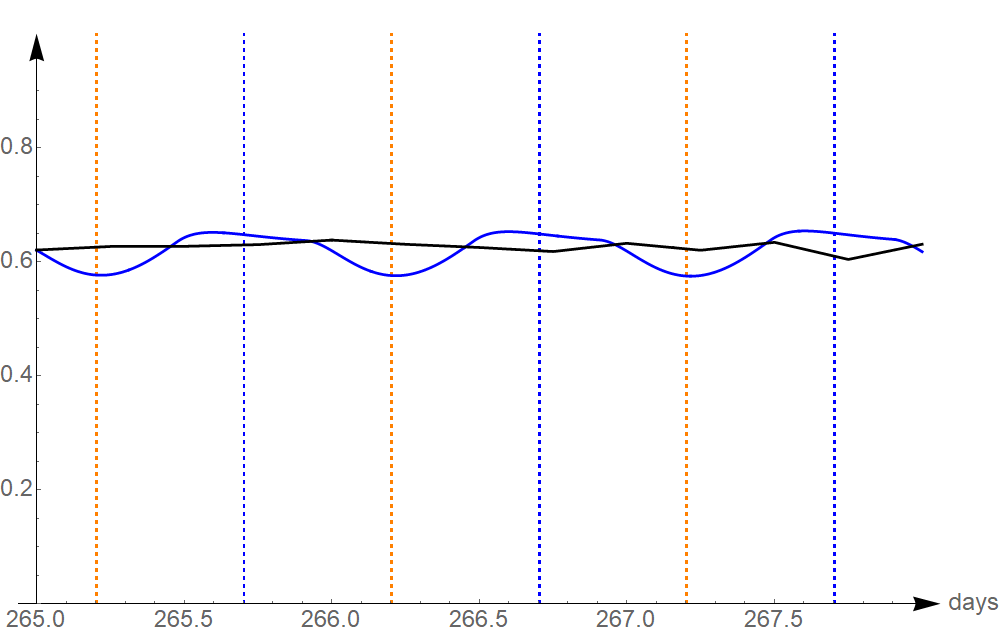}\includegraphics[width = 0.25 \textwidth]{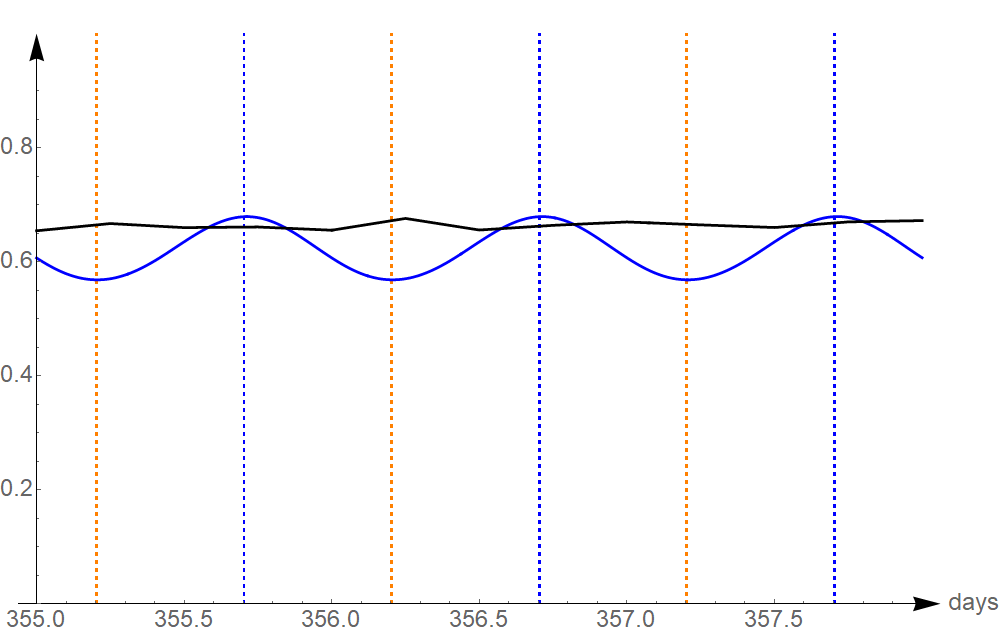}
\caption{Mean and computed temperature (left) and relative humidity (right) in Vostok during the year (top) and on solstices and equinoxes (bottom). In black the average temperatures and humidities, in red simulated temperatures, in blue simulated humidity, the grid lines represent solstices, equinoxes, noons and midnights.}\label{fig:vostok}
\end{center}
\end{figure}

In Figure~\ref{fig:vostok} are represented the computed evolution of temperature (red) and humidity (blue) of the air and the real averaged temperatures and humidities (black) from 1973 to 2019.

\subsection{More eccentric cases}

\begin{figure}[h]
	\begin{center}
\includegraphics[width = 0.5\textwidth]{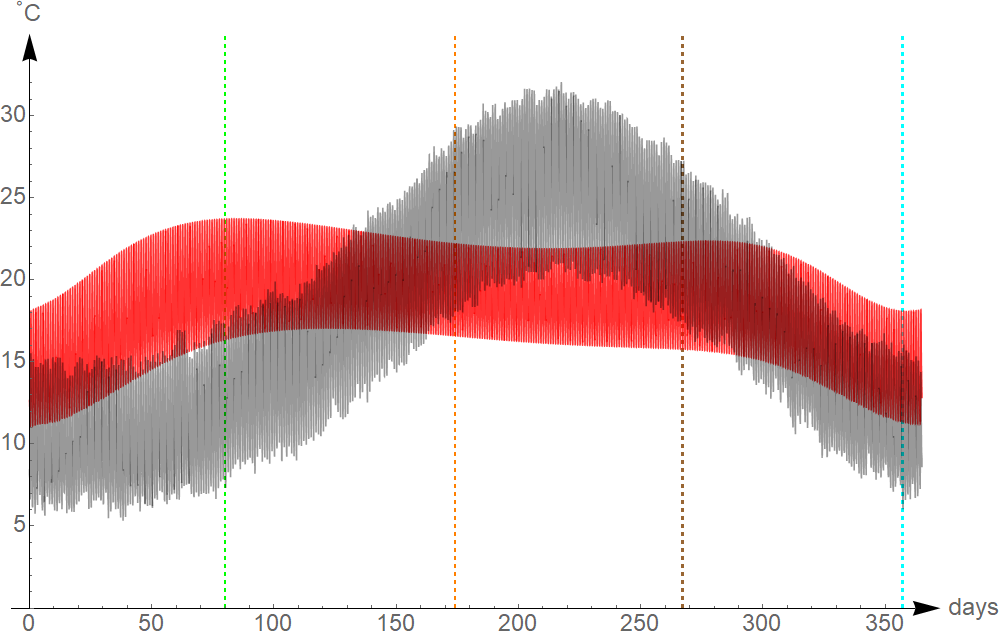}\includegraphics[width = 0.5\textwidth]{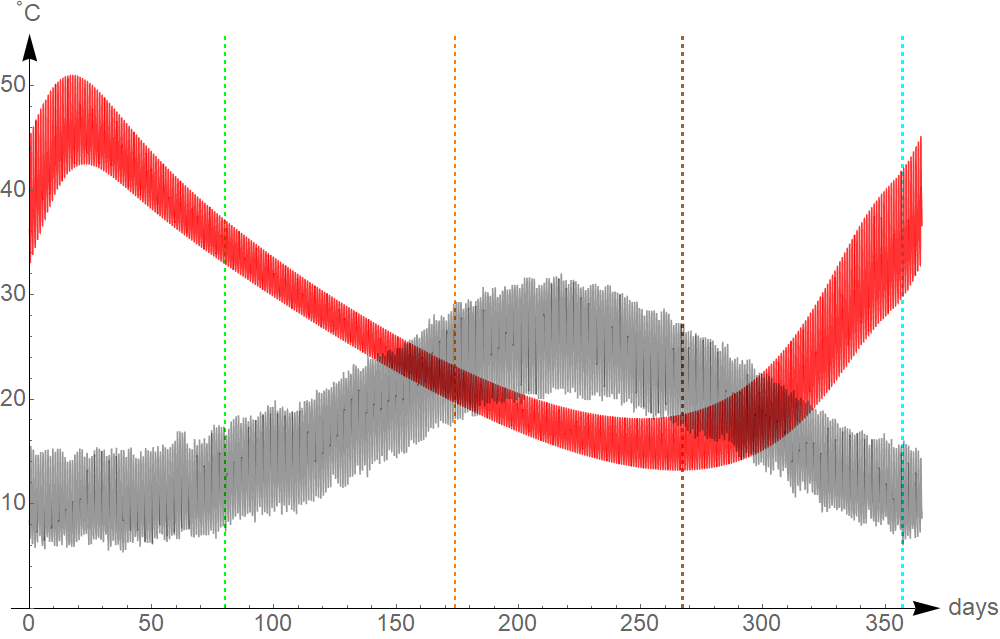}
\includegraphics[width = 0.5\textwidth]{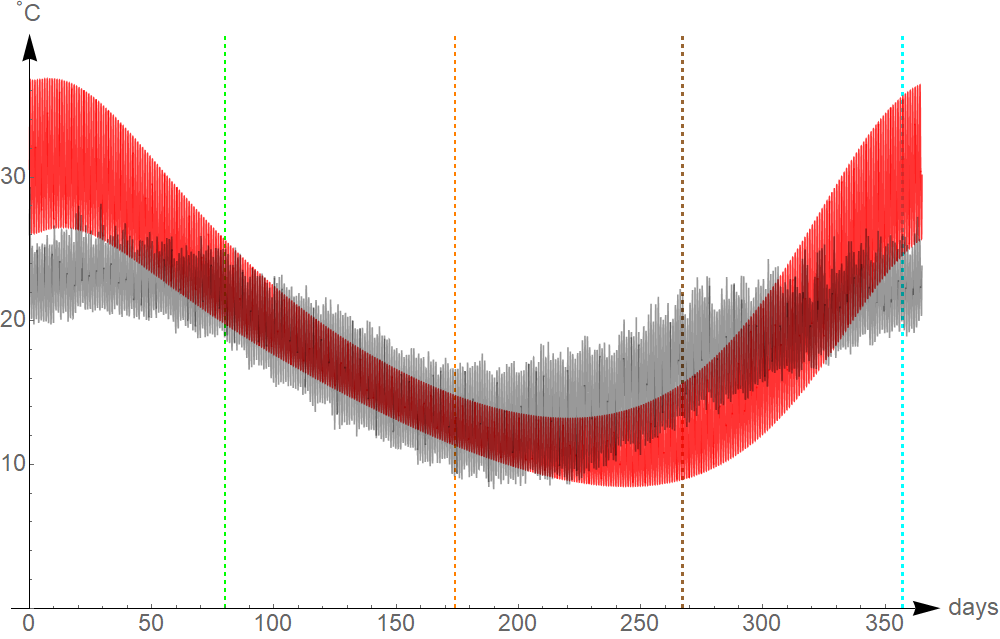}\includegraphics[width = 0.5\textwidth]{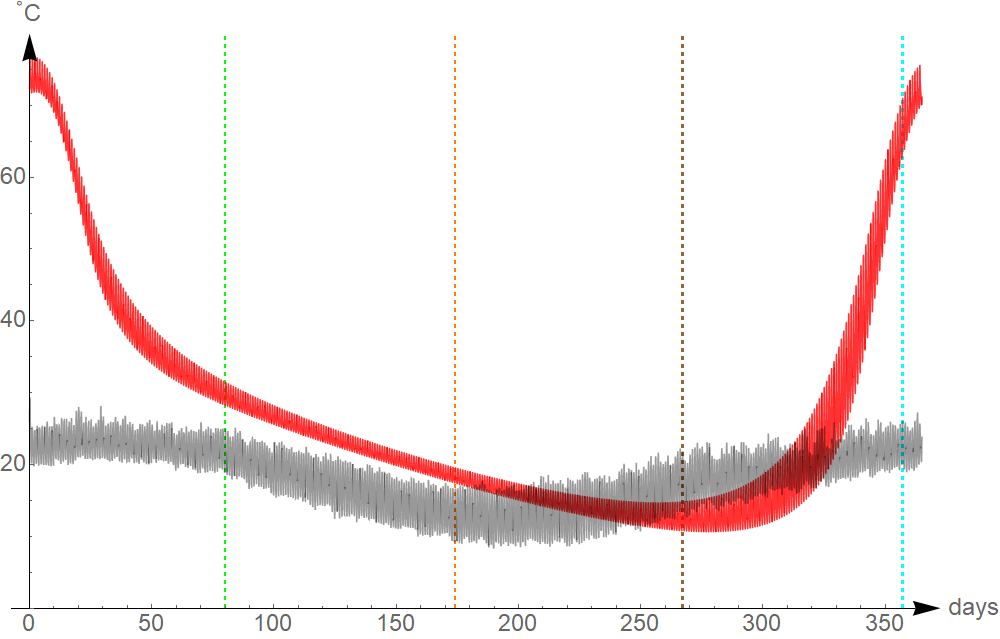}
\caption{Temperatures in Catania (above) and in Sydney (below) if the eccentricity of Earth's orbit was 0.2 (left) or 0.5 (right). In black the real temperatures, in red the ones obtained with the model.}
\label{fig:eccentrico}
	\end{center}
\end{figure}

It is well known that the for solar system the major semiaxis of the Earth's orbit is stable under perturbations while the stability of the full set of orbital parameters is still much discussed in modern times \cite{2019.SP.Laskar}. In our model the orbital parameters can be easily changed to model the temperatures in an Earth-like planet. The small eccentricity of the orbits in the solar system are well known to be non-generic \cite{2004.NA.Gaidos.Williams}. In the following plots we investigate the temperatures that Catania would have if the eccentricity of Earth was $e = 0.2$ or $e = 0.5$, and we compare the same effect on Sydney, a city in the southern hemisphere. We recall that, because of Earth's orientation of the rotation axis, during the summer of the northern hemisphere the Earth is at the aphelion, while during the summer of the southern hemisphere the Earth is at the perihelion. It follows that the effect of a change in eccentricity is mild in Catania (see Figure~\ref{fig:eccentrico} top) and severe in Sydney (see Figure~\ref{fig:eccentrico} bottom). Let us note however that the precession of the equinoxes would switch the situation every half Platonic year (12886 years).

These speculations are particularly interesting for their applications on exoplanets, where the suitability of temperatures to host life is a fundamental issue \cite{Demory2016, 1995.JC.Thompson.Pollard}.

\section{Conclusions}\label{conclusioni}

In this article, starting from basic physical laws, observations on the geometry of solar systems, and knowledge of the structure of a given planet, we design a model for local climate. The parameters involved in this dynamical model are mostly given by experimental experience. Our investigation is restricted to Earth, for which we can reproduce climatic phenomena like lag of seasons, lag of noons, asymmetry of daily temperatures, evolution of temperatures related a variety of K\"oppen climatic zones.

The temperatures computed solving the equations are reasonably similar to the real ones. In particular, as the real one, display lags and asymmetric evolution. They could be better fitted with a more detailed choice of the parameters or adding some other phenomenon to the equations. The annual excursion is very reasonable in all models, the daily temperature excursions tend to be slightly smaller than real (see Figures~\ref{fig:kufra}--\ref{fig:lincoln}). Despite the fact that humidity creates an asymmetric raise and fall of temperatures, the raise of real temperatures in the morning is faster than simulated ones. Simulated humidities tend to be constant during the year, while in some regions real humidities have smaller values in the summer than in the winter; the daily excursions are much more accurate.

The model still displays some criticalities and it can be improved in many ways. In particular we indicate the following issues:
\begin{enumerate}
\item the model requires the inclusion of some water also when dealing with desert or ice-caps, because the stabilising effect of water is necessary to avoid high annual temperature excursions;
\item we only consider the lowest part of the atmosphere.
\item spatial diffusion has been disregarded;
\item the equations that model humidity is not completely satisfactory, it probably should take into account other factors; 
\end{enumerate}

The first and second issues could be dealt with by adding other layers, one below the soil and one above the lower atmosphere. This would grant a correct annual excursion without compromising the daily one.

The spatial diffusion has been intentionally excluded to keep the model as simple as possible. The introduction of diffusion completely changes the approach, forcing a discretisation of the surface of the planet and the creation of a GCM which requires a detailed description of the planet surface and a large computational effort.

The evaporation depends on wind velocity, and probably non-constant wind speed should be taken into account, as well as seasonal rainfall ratio variation. We have not made a deep investigation on this facts, and we do not propose solutions.

The investigation is suitable to applications to exoplanets. Some astronomical parameters of exoplanets are known, but the choice of most other parameters is a delicate issue and will be subject for future works. In particular we think that the model could be most useful in the investigation of habitable and tidally locked planets.

\subsection*{Acknowledgements}

AG wishes to thank Giancarlo Benettin for his suggestion of considering two thermodynamic bodies in the model. GDB  thanks Alberto Chiavetta for his help in understanding the physics of the system. Both authors thank Paolo Falsaperla for enlightening discussions on energy balance and an anonymous referee for his critical reading of a first version of the article. 

AG is supported by the group GNFM of INdAM, by the grant No. PTRDMI-53722122113 of the University of Catania, and by Grant 2017YBKNCE of the national project PRIN of MIUR. GDB acknowledges support from Scuola Superiore di Catania.

%
%
%
%
%
%

\section{Appendix: the mathematical essence}\label{appendice}

In this appendix we make a cumbersome mathematical analysis of the fundamental reason that justifies the double-lag phenomenon. To model the temperature evolution of two thermodynamic bodies driven by a doubly-periodic forcing term, we consider a system of two differential equations:
\begin{equation}\label{sistema essenziale}
\begin{cases}
\dot Q_1 = - (a + c) Q_1 + d \, Q_2 + \\
\qquad \qquad + \delta_1\big[\alpha \sin (\omega\,t) \sin (\Omega  \, t) + \beta \cos (\omega \, t ) \cos (\Omega \, t) + \gamma \sin (\omega \, t)\big]\\[5pt]
\dot Q_2 = c \, Q_1 - (b + d) Q_2 +  \\
\qquad \qquad +\delta_2\big[\alpha \sin (\omega \, t) \sin (\Omega \, t) + \beta \cos (\omega \, t) \cos (\Omega \, t) + \gamma \sin (\omega \, t)\big].
\end{cases}
\end{equation}
We have discussed in Section~\ref{fisica} how this system models the temperature evolution of two different thermodynamic bodies in a zone of a planet. The only difference with equations \eqref{2RCM} lays on the fact that the longitude is absent and the exchange of heat is not mediated by a layer of air. The two bodies are irradiated by solar rays modulated by two frequencies $\omega$ and $\Omega$ that are respectively $2\pi$ times the reciprocal of a year and $2 \pi$ times the reciprocal of a day. The terms $a \, Q_1$ and $b \, Q_2$ model the heat flow from the bodies to space, the terms $c \, Q_1$ and $d \, Q_2$ model the rate of heat exchange among the two bodies.

Using Prostaferesi-Werner formulaes one can rewrite the equations as
\begin{equation*}
\begin{cases}
\dot Q_1 = c Q_2 - (a + c) Q_1 +  \delta_1 \Big[ (\alpha+ \beta) \cos(\Omega_- t) + (\beta-\alpha) \cos(\Omega_+  t) + \gamma \sin (\omega \, t)\Big]\\[5pt]
\dot Q_2 = c Q_1 - (b + c) Q_2  + \delta_2 \Big[ (\alpha+ \beta) \cos(\Omega_- t) + (\beta-\alpha) \cos(\Omega_+ t) + \gamma \sin (\omega \, t)\Big],
\end{cases}
\end{equation*}
with $\Omega_- = \Omega - \omega$ and $\Omega_+ = \Omega + \omega$. The homogeneous linear system associated to the equations has a stable node for every choice of parameters. In fact the determinant of the associated matrix is
\[
(a+c)(a+d) - d c = a^2 + ad + c a > 0
\]
and the discriminant $tr^2 - 4 \, det$ is
\begin{multline*}
((a+c)+(b+d))^2 - 4 (a+c)(b+d) + 4 cd =  (a-b)^2 + (c+d)^2 + 2 (c-d)(a-b) \geq \\
\geq (a-b)^2 + (c-d)^2 + 2 (c-d)(a-b) = ((a-b) + (c-d))^2 \geq 0.
\end{multline*}


With a linear change of variables of matrix $(S_1,S_2) = P(Q_1,Q_2)$ the system becomes
\begin{equation*}
\begin{cases}
\dot S_1 = - \lambda_1 S_1  + \chi_1 \big[ (\alpha+ \beta) \cos(\Omega_- t) + (\beta-\alpha) \cos(\Omega_+ t) + \gamma \sin (\omega \, t)\big]\\
\dot S_2 = - \lambda_2 S_2 + \chi_2 \big[(\alpha+ \beta) \cos(\Omega_- t) + (\beta-\alpha) \cos(\Omega_+ t) + \gamma \sin (\omega \, t)\big],
\end{cases}
\end{equation*}
where the vector $(\chi_1, \chi_2) = P (\delta_1,\delta_2)$, $P$ is the matrix of change of basis, $S = P Q$, and $\lambda_1,\lambda_2$ are the two eigenvalues of the linear system. The actual expression of the coefficients $\lambda_1$, $\lambda_2$, $\chi_1$, $\chi_2$ is irrelevant for our purposes. What is important is that the asymptotic solutions to these equations have the form 
\begin{equation*}
\begin{cases}
S_1 = \chi_1 \left[ (\beta + \alpha) \frac{\cos\left(\Omega_- (t - \tau_1^-)\right)}{\sqrt{\lambda_1^2 + \Omega_-^2}}  + (\beta - \alpha) \frac{\cos\left(\Omega_+ (t - \tau_1^+)\right) }{\sqrt{\lambda_1^2 + \Omega_+^2}} + \gamma \frac{\sin\left(\omega (t - \tau_1)\right)}{\sqrt{\lambda_1^2 + \omega^2}}  \right]\\[10pt]
S_2 = \chi_2 \left[(\beta + \alpha) \frac{\cos\left(\Omega_-(t - \tau_2^-)\right)}{\sqrt{\lambda_2^2 + \Omega_-^2}}  + (\beta - \alpha) \frac{\cos\left(\Omega_+ (t - \tau_2^+)\right)}{\sqrt{\lambda_2^2 + \Omega_-^2}}  + \gamma \frac{ \sin\left(\omega (t - \tau_2)\right)}{\sqrt{\lambda_1^2 + \omega}} \right],
\end{cases}
\end{equation*}
with
\[
\tau_i^\pm = \varphi_i^\pm/\Omega_\pm, \quad \varphi_i^\pm  = \arg(\lambda_i + i \Omega_\pm), \quad \tau_i = \varphi_i/\omega, \quad \varphi_i= \arg(\lambda_i + i \omega)
\]
for $i= 1,2$. Turning back to the temperatures $T_1,T_2$ one has
\[
{\small \begin{pmatrix} Q_1\\Q_2 \end{pmatrix} = 
P^{-1} 
\begin{pmatrix}
\chi_1 \left[ (\beta + \alpha) \frac{\cos\left(\Omega_- (t - \tau_1^-)\right)}{\sqrt{\lambda_1^2 + \Omega_-^2}}  + (\beta - \alpha) \frac{\cos\left(\Omega_+ (t - \tau_1^+)\right)}{\sqrt{\lambda_1^2 + \Omega_+^2}} + \gamma \frac{\sin\left(\omega (t - \tau_1)\right)}{\sqrt{\lambda_1^2 + \omega^2}}  \right]\\[10pt] 
\chi_2 \left[(\beta + \alpha) \frac{\cos\left(\Omega_-(t - \tau_2^-)\right)}{\sqrt{\lambda_2^2 + \Omega_-^2}}  + (\beta - \alpha) \frac{\cos\left(\Omega_+ (t - \tau_2^+)\right)}{\sqrt{\lambda_2^2 + \Omega_-^2}}  + \gamma \frac{\sin\left(\omega (t - \tau_2)\right) }{\sqrt{\lambda_2^2 + \omega^2}} \right].
\end{pmatrix}}
\]
The term
\[
\gamma P^{-1}  \begin{pmatrix}  \frac{\chi_1}{\sqrt{\lambda_1^2 + \omega^2}}  \sin\left(\omega (t - \tau_1)\right) \\   \frac{\chi_2}{\sqrt{\lambda_2^2 + \omega^2}} \sin\left(\omega (t - \tau_2)\right)
\end{pmatrix} 
\]
is responsible of the yearly delay, that can be estimated with the following algebraic steps:
\begin{multline*}
\gamma P^{-1}  \begin{pmatrix}  \frac{\chi_1}{\sqrt{\lambda_1^2 + \omega^2}}  \sin\left(\omega (t - \tau_1)\right) \\   \frac{\chi_2}{\sqrt{\lambda_2^2 + \omega^2}} \sin\left(\omega (t - \tau_2)\right) \end{pmatrix} = \gamma P^{-1}  \begin{pmatrix}  \frac{\chi_1}{\lambda_1^2 + \omega^2}  (\lambda_1 \sin(\omega t) - \omega \cos(\omega \, t)) \\   \frac{\chi_2}{\lambda_2^2 + \omega^2} (\lambda_2 \sin(\omega t) - \omega \cos(\omega \, t)) \end{pmatrix}
\\[3pt]
= \gamma  \begin{pmatrix}  
\frac{\chi_{11} \chi_1}{\lambda_1^2 + \omega^2}  (\lambda_1 \sin(\omega t) - \omega \cos(\omega \, t)) + \frac{\chi_{12} \chi_2}{\lambda_2^2 + \omega^2} (\lambda_2 \sin(\omega t) - \omega \cos(\omega \, t)) \\  \frac{\chi_{21} \chi_1}{\lambda_1^2 + \omega^2}  (\lambda_1 \sin(\omega t) - \omega \cos(\omega \, t)) + \frac{\chi_{22} \chi_2}{\lambda_2^2 + \omega^2} (\lambda_2 \sin(\omega t) - \omega \cos(\omega \, t)
\end{pmatrix}
\\[3pt]
= \gamma  
\begin{pmatrix}  
\left(\frac{\chi_{11} \chi_1 \lambda_1}{\lambda_1^2 + \omega^2} + \frac{\chi_{12} \chi_2 \lambda_2}{\lambda_2^2 + \omega^2}\right) \sin(\omega \, t) - \omega \left(\frac{\chi_{11} \chi_1}{\lambda_1^2 + \omega^2} + \frac{\chi_{12} \chi_2}{\lambda_2^2 + \omega^2} \right) \cos(\omega \, t) \\  
\left(\frac{\chi_{21} \chi_1 \lambda_1}{\lambda_1^2 + \omega^2} + \frac{\chi_{22} \chi_2 \lambda_2}{\lambda_2^2 + \omega^2} \right)\sin(\omega \, t) - \omega \left(\frac{\chi_{21} \chi_1}{\lambda_1^2 + \omega^2} + \frac{\chi_{22} \chi_2}{\lambda_2^2 + \omega^2} \right) \cos(\omega \, t)
\end{pmatrix}.
\end{multline*}

The lag of seasons for the bodies 1 and 2 are the two components of the vector
\begin{multline*}
\begin{pmatrix}
\arg\left[ \left(\frac{\chi_{11} \chi_1 \lambda_1}{\lambda_1^2 + \omega^2} + \frac{\chi_{12} \chi_2 \lambda_2}{\lambda_2^2 + \omega^2}\right) + i \omega \left(\frac{\chi_{11} \chi_1}{\lambda_1^2 + \omega^2} + \frac{\chi_{12} \chi_2}{\lambda_2^2 + \omega^2} \right)\right]\\[10pt]
\arg\left[\left(\frac{\chi_{21} \chi_1 \lambda_1}{\lambda_1^2 + \omega^2} + \frac{\chi_{22} \chi_2 \lambda_2}{\lambda_2^2 + \omega^2}\right) + i \omega \left(\frac{\chi_{21} \chi_1}{\lambda_1^2 + \omega^2} + \frac{\chi_{22} \chi_2}{\lambda_2^2 + \omega^2} \right)\right]
\end{pmatrix} = \\
= \arg\left[P^{-1} \begin{pmatrix} \frac{\lambda_1 + i \omega}{\lambda_1^2 + \omega^2} & 0 \\ 0 & \frac{\lambda_2 + i \omega}{\lambda_2^2 + \omega^2} \end{pmatrix} P \begin{pmatrix} \delta_1 \\ \delta_2 \end{pmatrix} \right] = \begin{pmatrix} \sigma_1 \\ \sigma_2 \end{pmatrix}.
\end{multline*}

The lag of noon is more delicate. In fact the delay can be estimated only if the ratio $\Omega/\omega$ is large (as it happens on Earth). In such case the evolution of temperatures is the sum of two terms:
\[
P^{-1} \left[ (\alpha + \beta)  \begin{pmatrix}  \chi_1  \frac{\cos\left(\Omega_- (t - \tau_1^-)\right) }{\sqrt{\lambda_1^2 + \Omega_-^2}}  \\  \chi_2 \frac{\cos\left(\Omega_- (t - \tau_2^-)\right)}{\sqrt{\lambda_2^2 + \Omega_-^2}} 
\end{pmatrix}  +
(\beta-\alpha)  \begin{pmatrix}  \chi_1 \frac{\cos\left(\Omega_+ (t - \tau_1^+)\right)}{\sqrt{\lambda_1^2 + \Omega_+^2}}   \\  \chi_2  \frac{\cos\left(\Omega_+(t - \tau_2^+)\right)}{\sqrt{\lambda_2^2 + \Omega_+^2}} 
\end{pmatrix}  \right].
\]
If the ratio $\Omega/\omega$ is large, then
\[
\tau_i^- \simeq \tau_i^+ \simeq \frac{\arg(\lambda_i + i \Omega)}{\Omega} : = \zeta_i, \qquad \sqrt{\lambda_i^2 + \Omega_\pm^2} \simeq \sqrt{\lambda_i^2 + \Omega^2}.
\]
It follows that the lag of noon of the two bodies is given by
\[
\arg \left[P^{-1} \begin{pmatrix} \frac{\lambda_1 + i \Omega}{\lambda_1^2 + \Omega^2} & 0 \\ 0 & \frac{\lambda_2 + i \Omega}{\lambda_2^2 + \Omega^2} \end{pmatrix} P \begin{pmatrix} \delta_1 \\ \delta_2 \end{pmatrix} \right] = \begin{pmatrix} \nu_1 \\ \nu_2 \end{pmatrix}.
\]

More precisely, one has that the solutions to the equation \eqref{sistema essenziale} are
\[{\small
\begin{pmatrix} Q_1\\Q_2 \end{pmatrix} \simeq 
\begin{pmatrix}
\hat \delta_1 \big[\alpha \sin (\omega\,(t-\nu_1)) \sin (\Omega  \, (t-\nu_1)) + \beta \cos (\omega \, (t-\nu_1)) \cos (\Omega \, (t-\nu_1)) \big] + \\[2pt]
\hskip -6cm +\hat \gamma_1 \sin (\omega \, (t- \sigma_1))
\\[8pt]
\hat \delta_2 \big[\alpha \sin (\omega\,(t-\nu_2)) \sin (\Omega  \, (t-\nu_2)) + \beta \cos (\omega \, (t-\nu_2)) \cos (\Omega \, (t-\nu_2)) \big] +\\[2pt]
\hskip -6cm + \hat \gamma_2 \sin (\omega \, (t- \sigma_2))
\end{pmatrix}.}
\]


\bibliography{library.bib}{}
\bibliographystyle{plain}
\end{document}